\documentstyle[12pt]{article}
\title{Heterogeneous condensation on the centers with
continuous activity in dynamic conditions}
\author{V.Kurasov}

\date{Victor.Kurasov@pobox.spbu.ru}

\begin{document}
\maketitle
\begin{abstract}

A system with  a metastable phase and a
pseudo continuous set  of the heterogeneous centers
is
considered.
An analytical theory for kinetics of the process of condensation
in such a system is constructed.
The free energy of formation of the  critical embryo
 is assumed to be known in
the capillary (macroscopic) approach as well as the solvatation energy
of the embryo.
The theory is
based on the quasistationary approximation for the nucleation rate which
has been  justified analytically.
An effective iteration procedure is presented. The iterationa are calculated
analytically. The approximate universal form of the spectrum is established.
\end{abstract}

\pagebreak

\section{Introduction}

The theory considered here
completes the program of construction of the theory of the aerosol formation
announced in \cite{aero1}. It will be based on the capillary approximation of
the height of the activation barrier.
This approximation lies in the base of the classical theory of
the homogeneous nucleation.
All necessary bibliographic remarks can be found in \cite{aero1}.

 Speaking about the activity
of the heterogeneous centers we mean that the different activity initiates
the different height of the activation barrier $\Delta F$, i.e. the
difference between the free energy of the critical embryo $F_{c}$ and
the free energy of solvatation $G$. The set
of the different activities of the heterogeneous centers can be
so dense that we can regard it as the pseudo continuous
one. In the case of
the "solid nucleus of condensation with the weak interaction"
the continuous size of the nucleus ensures the continuous set of $\Delta F$.

The ordinary external conditions for the phase transition have  the smooth
character in time. The external  action on the system leads to formation
of the droplets  of the new phase. The process of condensation violates
the thermodynamic parameters of the system. When the external action on
these parameters has the smooth behavior in time we shall say that
condensation occurs under the dynamic conditions.

Nevertheless the theory of the heterogeneous condensation in the dynamic
conditions
 has been  constructed only
for one type of the heterogeneous centers. So, the task to construct the
 kinetic theory
 for the system with a continuous set of the heterogeneous centers is rather
essential. It will be completely fulfilled here.

We shall use the physical assumptions
analogous to \cite{aero1}
which are necessary to construct the mathematical model:
\begin{itemize}
\item
the thermodynamic description of the critical embryo,
\item
the random homogeneous space distribution of the
heterogeneous centers,
\item
the free-molecular regime of the droplets growth,
\item
the homogeneous
external conditions for the   temperature and for the pressure,
\item
rather a high
activation barrier.
\end{itemize}
As far as the most interesting characteristics of this process are the
 numbers of the heterogeneously
formed droplets on the centers with a different activity
we shall estimate the accuracy of the theory
by
 the error of the obtained solutions for
 these values\footnote{But not for the total number of the droplets.}.
The unit volume is considered.
All energy-like values are measured in the thermal units.

The publication has the following structure:
\begin{itemize}
\item
In the first       part the system of the equations of condensation
is constructed.
\item
In the second part the explicit calculation of the iterations is presented.
\item
In the third part the conception of the universal spectrum is developed.
\item
In the last part some realistic spectrum of activities is considered
and the principle  of the self-consistency of
the spectrum of the activities is presented.
\end{itemize}

We shall define the activity of the
heterogeneous center as some parameter $w$ which
is proportional to the height of the activation barrier
\begin{equation} \label{**}
\Delta F (w) = \Delta F \mid_{w=0} - \lambda w
\end{equation}
with some positive coefficient $\lambda$.
Note that the choice of the initial point $w=0$ is
rather arbitrary now. We suppose $\Delta F(w) \gg 1$ for all essential types
of the heterogeneous centers\footnote{When this condition is violated
the kinetics of the process can be described by the trivial modification
of the theory presented below.}.

The total number of the heterogeneous centers
with the given activity $w$ will be marked by $\eta_{tot}(w)$. Naturally
$\eta_{tot}(w)$ is rather a smooth function of $w$.
We shall suppose that $\eta_{tot}$ in the essential region is near some constant
value
or it can be well approximated by the polinom which power isn't too high.
This value is constant in time.

          The density
of the molecules in the equilibrium vapor is marked
by \( n_{\infty} \), the    density of the molecules in the
real vapor  is marked by \( n\).
 The power of the metastability will be characterized by the
value of the supersaturation
$$ \zeta = \frac{ n - n_{\infty} }{ n_{\infty} } $$
We shall define the super-critical embryos as the "droplets".
Every droplet is described by the number of the molecules
\( \nu \) , or by the linear size $$ \rho = \nu^{1/3} $$
Due to the free-molecular regime of the droplets growth we have
$$ \frac{d\rho}{dt} = \zeta \alpha \tau^{-1} $$
where \( \alpha \) is the condensation coefficient and \( \tau \) is
some characteristic  time between
the collisions in the saturated vapor obtained from the gas kinetic theory.

Let us introduce the  size \( z \)
according to
\begin{equation}
\label{2}
z = \int_{t_*}^{t} \zeta \alpha \tau^{-1} dt'
\end{equation}
Here $t_*$ is some characteristic moment of time which belongs to the
period of the intensive formation of the droplets. The choice of $t_*$
is rather arbitrary. One can use for example the choice
 described in \cite{Novosib}.
Until the beginning of the coalescence \cite{15},\cite{16}
which isn't considered here
equation (\ref{2}) ensures the growth of \( z \) in time and can be
inverted
\begin{equation}
t(z) = \int_{0}^{z}  \tau \alpha^{-1} \frac{dx}{\zeta(x)}
+ t_*
\end{equation}
Hence, all values dependent on time become the values dependent on
\( z\) and the relative size $$ x=z-\rho $$ can be introduced.
During the whole evolution the droplet has one and the same
value of the variable \( x \).
Considering \( t(x) \) as the moment when the droplet with the given $x$ has
been formed (as a droplet) we can consider all functions of time as
the functions
of \( x \) .
Hence, we can see that the kinetic equation is reduced to the
fact that every droplet keeps the constant value of $x$. To reconstruct the
picture
of the evolution one must establish the dependencies $t(z)$ and $\zeta(x)$.

The values at the moment $t_*$ will be marked  by the lower index "*".
The positions of the region of the intensive formation of the droplets
are essentially  different. But one can  introduce $t_*$ as the moment
corresponding  to the maximum af the supersaturation.

\section{The system of the equations of condensation}

We shall mark by the argument \( \infty \) the total values of the magnitudes
formed during the whole condensation process.

Introduce the value of $$\zeta_{ideal} = \frac{n_{tot}}{n_{\infty}} -1
$$  where $n_{tot}$ is the total number of the molecules in the system.

We must take into account the reduction of $\zeta_{ideal}$ to some value
$\Phi$ due to the consumption of the vapor molecules in the process of
solvatation
\cite{Sevdec} \cite{Specdec} \cite{Sevdin}. Moreover, according to
\cite{Sevdin} the ideal supersaturation can be changed by
the external
supersaturation.
In any case we shall assume this value as the known one and mark it by
$\Phi$.
In some rough approximation $\zeta_{ideal} \approx  \Phi$.

We shall mark by $\eta^{tot}$  the total number of
the heterogeneous centers of
all
types:
\begin{equation}
\eta^{tot}= \int dw \eta_{tot}(w)
\end{equation}
where $\eta_{tot}(w)$ is the total number of the heterogeneous centers of
the
given type (the density of the total value of the heterogeneous centers).

The following statements are valid in the further consideration:
\begin{itemize}

\item
(1) The main role in the vapor consumption during the evolution
is played by the super-critical embryos, i.e. by the droplets.

\item
(2) The quasistationary approximation for the nucleation rate is valid during the
period of
the essential formation of the droplets for those sorts of the heterogeneous
centers which aren't completely exhausted in this process.

\end{itemize}

The justification of the second statement\footnote{The second statement
isn't valid for those sorts of the heterogeneous centers which are going
to be completely exhausted. But there the result is obvious. Note  that
the periods of the intensive formation of the droplets on the centers
with the different activities don't coincide.} uses the estimate for times
 \( t^{s}_{i} \)
of the relaxation to  the stationary  state
in the near-critical region  which can be found in \cite{3},
\cite{17}
(for the investigation of the
heterogeneous barrier  the consideration is  the same one).

Let \( f_{s} \) be the stationary value of
the distribution of the sizes of the heterogeneously formed droplets
 measured in the units of \( n_{\infty} \).
It can be presented in the following form
\begin{equation}
f_{s} = f_{\zeta}(\zeta (x),w) \eta(x,w)
\end{equation}
where $\eta(x,w)$ is the density on activities of the number of
the heterogeneous
centers which are
free from the super-critical embryos and $f_{\zeta}$ is given by the following
formula \cite{18}
\begin{equation}
f_{\zeta}=\frac
{
 W^{+}_{c} \exp(-\Delta F(\zeta,w)) \tau
}
{
 \pi^{1/2} \Delta_{e} \nu  \Delta_{c} \nu  \zeta \alpha  n_{\infty}
}
\end{equation}
where $W^{+}$ is the number of the molecules in the vapor
which interact with the droplet in the unit
of time, $\Delta_{e} \nu$  is the width of the equilibrium distribution
$$
\Delta_{e} \nu = \sum_{\nu=1}^{\nu=(\nu_{c}+\nu_{e})/2}\exp(-F(\nu)+G)
$$
and $\Delta_{c} \nu$ is the halfwidth of the near-critical region
$$
\Delta_{c} \nu =
\frac{2^{1/2}}{\mid
(\frac{\partial^{2} F }{\partial \nu^{2}})_{\nu=\nu_{c}} \mid^{1/2} }
$$
Index "$c$" marks the values for the critical embryo and "e" - the values
for the equilibrium embryo.
Certainly, $\Delta_c \nu$ and $\Delta_e \nu$ are some smooth functions
of $w$ and we shall neglect this dependence.

We shall mark by \( n_{\infty} g(w) \) the density of the distribution on activities
of the total number of the molecules of the condensated substance
 in the heterogeneous  droplets
formed on the centers of the activity $w$.
To simplify the formulas we shall use $$ \theta(w) =
{\eta(w)}/{\eta_{tot}(w) } $$

 We obtain for \( g_{i},  \theta_{i} \) the following
equations
\begin{equation}
g(z,w) =   \int_{-\infty}^{z} (z-x)^{3} f_{\zeta}(\zeta(x),w) \eta(w) dx
\end{equation}
\begin{equation}
\theta(z,w) = \exp ( - n_{\infty} \int_{-\infty}^{z} f_{\zeta}(\zeta(x),w) dx )
\end{equation}

As far as we measure the accuracy of the theory in the terms of
the error in the droplets number
 we  define these values as the following ones:
\begin{equation}
N(z,w) = \eta_{tot}(w) ( 1 - \theta(z,w))
\end{equation}
The total number of the droplets is
\begin{equation}
N^{tot} = \int \eta_{tot}(w)( 1 - \theta(z,w)) dw
 = \int N(z,w) dw
\end{equation}

For the majority of the types of the heterogeneous centers
 the following approximations  of the nucleation rates
 are valid during the period of the essential formation of the droplets\footnote{Note
that the position of the regions of the intensive formation of the droplets
on some sort essentially depends on the activity of the centers. the following
approximation isn't valid for the regions of the intensive formation on
the active centers. But for these centers the result is evident - all
centers are going to become the centers of the droplets.}
\begin{equation}\label{expap}
f_{\zeta}(\zeta(x),w) =
f_{\zeta}(\Phi_{*},w) \mid_{w=0}
\exp ( \Gamma \frac{ ( \zeta - \Phi_{*} ) }
{ \Phi_{*} } )
 \exp( w \lambda )
\equiv
f_{\zeta\ *}
\exp ( \Gamma \frac{ ( \zeta - \Phi_{*} ) }
{ \Phi_{*} } )
 \exp( w \lambda )
\end{equation}
where
\begin{equation} \label{8}
\Gamma = -\Phi_{*}
\frac{d \Delta F(\zeta)}{d \zeta }  \mid_{\zeta=\Phi_{*}} \mid_{w=0}
\end{equation}
$$          f_{\zeta\ *}        = f_{\zeta}(\Phi_{*},w) \mid_{w=0}$$
and \( \Delta F \) is the height of the heterogeneous
activation barrier\footnote{The validity of these approximations may
fail in the situations of the extremely long spectrums (see later). But
the square form of the behavior of the supersaturation near the maximum will
be attained earlier.}.
The validity of these approximations can be  justified
for the heterogeneous embryos with
the interaction between the center and the molecules of the condensated phase
 weaker or equal than the reciprocal to the space distance.
So, we can imagine this as a hard sphere with a weak interaction  on which
the embryo  is formed.

The dependence of $\Gamma$ on $w$ is rather weak. So we can put
\begin{equation}
\Gamma(w) = \Gamma  \mid_{w=0}
\end{equation}
for any essential part of the spectrum of the activities. The applicability of the
last approximation is based on the following
qualitative model. It is known that $\Gamma$
is approximately
$\Phi_* (\nu_c - \nu_e) / (\Phi_* + 1) $.
When the supersaturation is sligtly changed then the value
of $\nu_e$ doesn't undergo some big variations and the variation of $\Gamma$ occurs
mainly due to the variation of $\nu_c$. The behavior of the value  $\nu_c$
resembles
the analogous behavior in the homogeneous case. As a result we can justify not
only the approximation concerning the dependence on the supersaturation,
but also the dependence on the activity.

 A natural question about  the essential part of the spectrum
appears here.
The process of condensation on the centers with some relatively high activity
occurs earlier than the
supersaturation attains maximum.  These centers form the droplets which
can be included in the value of the external supersaturation in the manner
of \cite{Sevdin} \cite{aero3}. For these centers the number of
the formed droplets
coincides approximately
with the number of the heterogeneous centers - all centers are
now the centers of
the droplets.

The action of this part of the spectrum on the further evolution can be treated
in the terms of the external supersaturation $\Omega$ (see \cite{Sevdin}).
The nontrivial statement that the process of formation of the droplets
on the centers with the intermediate activity occurs near the maximum of
the supersaturation
lies in the base of such a method
of description. This fact can be analytically proved.

The centers with some relatively low activity remain practically
unexhausted (when  the spectrum $\eta_{tot}$ has no singularities they can
not play any important role in condensation). The
singularities
can be described in this region in a manner from \cite{Sevdin} \cite{aero3}.
So only some  centers with the intermediate activity
are essential near the maximum of
the supersaturation.
In the scale of the activities this region corresponds to the variation of $w$
by the value of the order of $\lambda^{-1}$.

Now we shall formulate the system of the balance equations for the functions
$g(x,w)$, $\theta(x,w)$ and
$$ G(x) = \int dw g(x,w)$$
where the integral
is taken over the whole spectrum of the heterogeneous centers. In principle
we can write that  the region of integration goes from $-\infty$ up
to $\infty$ but in fact this integration must be carried out in the region
which covers the essential region of the spectrum\footnote{The extraction
of this region is quite analogous to the case of the decay on the spectrum
of activities.}. In this region it is necessary
to have relatively smooth  behavior of $\eta_{tot}$. This value must
be approximately constant
$$ \eta_{tot} \approx const $$
or must be well approximated by the polinom on the activities:
$$ \eta_{tot} = P_n (w) $$

We assume the  total number of the heterogeneous centers to be  constant in time.

Using the conservation laws for the heterogeneous centers
and for the molecules of the condensing
substance   we get for \( g,  \theta \) the following
equations
\begin{equation}
g(z,w) = f_{*}  \int_{-\infty}^{z} (z-x)^{3}
\exp ( -\Gamma \frac{ \zeta - \Phi_*  }
{ \Phi_{*} } )
\theta dx \exp( w \lambda )
\end{equation}
\begin{equation}
G(z)  = \int dw g(z,w)
\end{equation}
\begin{equation}
\theta(z,w)
 = \exp ( - \frac{f_{*}\exp(  \lambda w)  n_{\infty}}{\eta_{tot}}
\int_{-\infty}^{z}
\exp (  - \Gamma \frac{ \zeta - \Phi_*  }
{ \Phi_{*} } ) dx )
\end{equation}
\begin{equation}
\Phi = \zeta + G(z)
\end{equation}
where $f_{*} = f_{\zeta\ *} \eta_{tot}$

These equations form the closed system of the equations for condensation
kinetics. This system will be the subject of our investigation.

We shall consider this system of equations during the period when the
centers with the intermediate activity are going to become the centers of
the droplets. We shall call this period as the period of  the intensive formation
of the droplets (PIFD).

We assume that the ordinary \cite{Novosib}
 linearization of the ideal supersaturation
is valid during PIFD\footnote{One can  analytically show that the duration
if PIFD is rather short. Namely $$\frac{\Phi (z) - \Phi_*}{\Phi_*} \leq
\frac{1}{\Gamma}$$}
\begin{equation}
\Phi (x) =  \Phi_* + \frac{\Phi_*}{\Gamma} c x
\end{equation}
with some positive parameter $c$.

After the substitution of this linearization into the system of the condensation
equations this system transfers to
\begin{equation}
g(z,w) = f_{*}  \int_{-\infty}^{z} (z-x)^{3}
\exp (  cx - \frac{\Gamma}{\Phi_*} G(x) )
\theta(x,w) dx \exp( w \lambda )
\end{equation}
\begin{equation}
G(z)  = \int dw g(z,w)
\end{equation}
\begin{equation}
\theta(z,w)
 = \exp ( - \frac{f_{*}\exp(  \lambda w)  n_{\infty}}{\eta_{tot}}
\int_{-\infty}^{z}
\exp (  cx - \frac{\Gamma}{\Phi_*} G(x) ) dx )
\end{equation}

The spectrum of sizes can be found as the following one
\begin{equation}
f(x,w) = f_{*} \exp(\lambda w)
\exp ( -\Gamma \frac{\zeta - \Phi_* }
{ \Phi_{*} } )
\theta(x,w)
\end{equation}
and when the linearization is taken into account:
\begin{equation}
f(x,w) = f_{*} \exp(\lambda w)
\exp ( cx - \frac{\Gamma}{\Phi_*} G(x)
 )
\theta(x,w)
\end{equation}

\section{Iteration procedure}
Such systems as the already obtained one can be  solved
by the iteration procedure.
 It can be constructed by the following way: For the initial approximations
we choose:
\begin{equation}
g_{0}(z,w) = 0 \ \ \ \ \theta_{0}(z,w) = 1
\end{equation}
The recurrent procedure is defined according to
\begin{equation}
g_{i+1}(z,w) = f_{*}  \int_{-\infty}^{z} (z-x)^{3}
\exp (  cx - \frac{\Gamma}{\Phi_*} G_{i}(x) )
\theta_{i}(x,w) dx \exp( w \lambda )
\end{equation}
\begin{equation}
G_{i}(z)  = \int dw g_{i}(z,w)
\end{equation}
\begin{equation}
\theta_{i+1}(z,w)
 = \exp ( - \frac{f_{*}\exp(  \lambda w)  n_{\infty}}{\eta_{tot}}
\int_{-\infty}^{z}
\exp (  cx - \frac{\Gamma}{\Phi_*} G_{i}(x) ) dx )
\end{equation}

The chains of inequalities analogous to
\cite{PhysRev} guarantees the convergence of the
iterations
and some estimates analogous to \cite{PhysRev} can be established here.

The direct calculation of the iterations gives
\begin{equation}
g_{1}(z,w) = f_{*} \frac{6 \exp(cz)}{ c^4} \exp(\lambda w)
\end{equation}
\begin{equation}
\theta_{1}(x,w) = \exp(-f_{\zeta\ *} n_{\infty} \exp(\lambda w)
\frac{\exp(cx)}{c}
)
\end{equation}
\begin{equation}
G_1 (x) = f_{\zeta\ *} \frac{6 \exp(cx)}{ c^4} \int dw \exp(\lambda w)
\eta_{tot}
\end{equation}
The last integral causes some problems with convergence. When $\eta_{tot}
= const$ it can not be calculated. At least there are two possibilities
to overcome this difficulty.

The first possibility is  a more formal one. Certainly, in every system
$\eta_{tot}(w) = 0$ when $w$ is greater than some maximum value $w_{max}$.
So, formally the integral exists and it is equal to some constant. This
constant is generally unknown, but one can act in the manner like
it was done with the supersaturation in \cite{Novosib}. In \cite{Novosib}
the value of $\Phi_*$ was not the matter of consideration in the iteration
method, but was included in the set of the undefined parameters. After
the final form of the size spectrum was established an equation on parameters
was studied. The form of the spectrum can play a role of some ansatz with
several parameters. The same procedure can be inserted here. The reason
is that we can observe the separation of the functions containing $w$
from the terms containing the other parameters of the
condensation process. This fact
remains valid also in the second iteration which is rather a good
approximation for the final solution as it can be analytically proved
on the base of the analysis of the iteration procedure.

The second possibility seems to be more physical one because it allows
to combine the current problem with the problem of the correct definition
of the region of rather active and exhausted centers. Really we have not
 separated the region included into the external supersaturation from the
region which isn't included into the external supersaturation.
Note, that in the dynamic conditions the spectrum of sizes of the droplets
formed on some fixed sort of the heterogeneous centers has well defined boundaries.
The sizes (and the times of formation) of the droplets which are essential in
the vapor
consumption can be also well localized. Namely, the spectrum is  essentially
situated in
the region
\begin{equation}
                            - \frac{1}{c} \leq z \leq \frac{1}{c}
\end{equation}
and the region of the droplets essential for the vapor consumption during
PIFD is   essentially covered by the interval\footnote{See the analysis
of the second figure  from \cite{aero3}.}
\begin{equation}
- \frac{8}{c} \leq z \leq 0
\end{equation}
As a result one can state that for $z \sim -a/c$ where $a \sim 8$ one can
get the start of PIFD for the centers which are in the process of
nucleation
near the maximum of the supersaturation.

Note that the moment $t_*$ can be taken as a moment of the maximum of
the supersaturation. Really, when the maximum of intensity of formation (or
the moment of formation of
the half of the droplets) lies far from maximum of the supersaturation, then
the result of  nucleation of this sort of centers is rather obvious:
the supersaturation is equal to the ideal (or external) supersaturation
$$ \zeta = \Phi $$
the total number of the droplets is equal to the total  number of
the heterogeneous centers
$$ N_{tot} = \eta_{tot} $$
and the moment of formation of the spectrum on the centers of some fixed sort
which can be considered as
a monodisperse one can be found from
$$ \theta(z, w) = 1/2 $$
which is equivalent to
$$   f_{\zeta\ *} \exp(\lambda w) n_{\infty} \exp(cz) = c \ln 2 $$
where the first two terms can be changed by some slightly other amplitude of
the spectrum and the parameter $c$ can be reconsidered as far as the derivative
can vary.

The simplest regularization by the cut-off is the following one. We must
substitute
the value  of $\eta_{tot}$ by the value $\eta_{init}$ calculated as
$$ \eta_{init} = \eta_{tot} \theta(x=-a, w) $$
after the calculation of $\theta(-a,w)$ which can be calculated on the base
of the ideal supersaturation, i.e. in the first iteration. As far as
$$ \theta_{1}(-a,w) = \exp(- f_{\zeta\ *} n_{\infty} \exp(\lambda w)
\frac{\exp(-ca)}
{c})$$
the value of $f_*$ must be changed by
$$ f_* = f_{\zeta\ *} \eta_{tot}(w)
                      \exp(- f_{\zeta\ *} n_{\infty} \exp(\lambda w)
\frac{\exp(-ca)}
{c})$$

\section{Calculation of the iterations}

Let us calculate the iterations.
Then the expression for $G$ will be the following
$$ G_1 =
f_{\zeta\ *} \int_{-\infty}^{\infty} \eta_{tot}(w) \exp(\lambda w) \exp(-
A \exp(\lambda w)) dw \frac{6 \exp(cz)}{c^4}
$$
where
$$
A=
f_{\zeta\ *} n_{\infty} \frac{\exp(-ca)}{c}
$$
When $\eta_{tot} = const$ the integral can be taken which leads to
$$ G_1 = f_{\zeta\ *} \eta_{tot} \frac{1}{\lambda A} \frac{6 \exp(cz)}{c^4} $$
Note that the integral can be taken in the final limits which allows to
take into account only the finite region of activities as far as an
approximation
of the total number of heterogeneous centers by the composition of
the Heavisaid's
functions.

When $\eta_{tot}$ is approximated by the polinom it is easy to note that
each monom in the integral leads to some elementary functions with some
universal
constants. This fact can be seen from
the simple translation of the variable in $$
             \int_{-\infty}^{\infty} w^m \exp(\lambda w) \exp(-
A \exp(\lambda w)) dw  $$
to the variable $$\lambda w + \ln(A)$$
Here the integration must be fulfilled in the infinite
limits.
These integrals can be expressed through $\Psi$-function and
its derivatives when the argument is put to $1$.

The second approximation for $\theta$
leads to the following result:
$$ \theta_2(z,w) =
\exp(-f_{\zeta\ *} n_{\infty} \exp(\lambda w) \int_{-\infty}^{z}
\exp(cx - B \exp(cx) ) dx )
$$
where
$$
B = \frac{\Gamma}{\Phi_*} f_{\zeta\ *} \frac{6 \eta_{tot}}{c^4}
\frac{1}{\lambda
A}
$$
Note that the integral can be taken here also  in the finite limits.
After the integration we get
$$
\theta_{2}(z,w) =
\exp(-f_{\zeta\ *} n_{\infty} \exp(\lambda w) \frac{1}{c B} (1-\exp(-B\exp(cz))))
$$
Note that $\eta(w) = \eta_{init}(w) \theta(z,w)$.

Particularly, for the final values one can get
\begin{equation} \label{*}
 \theta_{2}(\infty, w) = \exp(- f_{\zeta\ *} n_{\infty} \exp(\lambda w)
\frac{1}{cB})
\end{equation}

This value of
$B$
is obtained in the approximation
$\eta=const$.
For the further consideration only the analytical structure of
$B$                                            is important.
Namely, in any case
(and also in the case
$\eta_{tot}(w) = P_n(w)$)
we have
for $G$ the expression with the following  analytical structure
$$
G_1 \sim const(z,w) \exp(cz)
$$
So,
$B$ is some constant and this fact ensures the possibility
of the further calculation of the iterations.

The next step is the calculation of the total number of the droplets  appeared
in the considered period which can be done by the simple integration
$$ N^{tot} = \int_{-\infty}^{\infty} dw N_{tot}(w) $$
Note that instead of
$$ N^{tot} = \int_{-\infty}^{\infty} dw \eta_{tot}(w)
(1 - \theta_{2}(\infty,w)) $$
which can not be integrated we must take
$$ N^{tot} =
\int_{-\infty}^{\infty} dw \eta_{init}(w) (1- \theta_{2}(\infty,w)) $$
which gives
           the integral with no problems of convergence.
To calculate the last integral note that
$$
\theta_2(-a,w) = \exp(
- f_{\zeta\ *}
\exp(\lambda w)
n_{\infty}
\int_{-\infty}^{-a} \exp(cx - \frac{\Gamma}{\Phi_*} G) dx
)
$$

Consider now the case $\eta_{tot} = const$.
After the evident renormalization of the variables of integration the integral
can be reduced to
$ \int_{-\infty}^{\infty}
\exp(-\exp(x)) (1-\exp(-H\exp(x))) dx$
where
$H$ is some constant. This value can be approximately calculated by the
following procedure.
Decompose the internal exponents and get
\begin{eqnarray}
\int_{-\infty}^{\infty}
\exp(-\exp(x)) (1-\exp(-H\exp(x))) dx = \nonumber \\
\int_{-\infty}^{\infty}
\frac{1}{
\sum_{i=0}^{\infty}
\frac{\exp(ix)}{i!}
}
(1-
\frac{1}{
\sum_{i=0}^{\infty}
\frac{\exp(ix)H^i}{i!}
}
)
dx    =  \\
\int_0^{\infty} \frac{
\sum_{i=0}^{\infty} \frac{y^i H^{i+1}}{(i+1)!}
}{
\sum_{i=0}^{\infty} \frac{y^i}{i!}
\sum_{i=0}^{\infty} \frac{y^i H^i}{i!}
}
dy
\nonumber
\end{eqnarray}
Note, that it is necessary to left the serial in the denominator.

The necessary accuracy will be ensured by the first three terms in
the decompositions of
$$
\int_0^{\infty} \frac{
\sum_{i=0}^{\infty} \frac{y^i H^{i+1}}{(i+1)!}
}{
\sum_{i=0}^{\infty} \frac{y^i}{i!}
\sum_{i=0}^{\infty} \frac{y^i H^i}{i!}
}
dy
\approx
\int_0^{\infty} \frac{
\sum_{i=0}^{2} \frac{y^i H^{i+1}}{(i+1)!}
}{
\sum_{i=0}^{3} \frac{y^i}{i!}
\sum_{i=0}^{3} \frac{y^i H^i}{i!}
}
dy
$$
Then
an integral from  the rational function can be simply calculated (also in
the finite limits).

Now  the integrals
$$
I_j \equiv \int_{0}^{\infty} \ln^j y
\frac{
\sum_{i=0}^{\infty} \frac{y^i H^{i+1}}{(i+1)!}
}{
\sum_{i=0}^{\infty} \frac{y^i}{i!}
\sum_{i=0}^{\infty} \frac{y^i H^i}{i!}
}
dy
\approx
           \int_{0}^{\infty} \ln^j y
\frac{
\sum_{i=0}^{n-1} \frac{y^i H^{i+1}}{(i+1)!}
}{
\sum_{i=0}^{n} \frac{y^i}{i!}
\sum_{i=0}^{n} \frac{y^i H^i}{i!}
}
dy
$$
will be calculated.
These integrals appear when the polynomial approximation for $\eta_{tot}$
is accepted.

Consider the function
$$ f(y) = \ln^{i+1}y
\frac{
\sum_{i=0}^{n-1} \frac{y^i H^{i+1}}{(i+1)!}
}{
\sum_{i=0}^{n} \frac{y^i}{i!}
\sum_{i=0}^{n} \frac{y^i H^i}{i!}
}
$$
as the function of a complex variable.
Integrate  this function along the closed path $\Omega$ constructed in
the following manner:
the big circumference with an infinite radius;
two straight lines
$]y+i0,\infty+i0[$, $]y-i0,\infty-i0[$;
the little circumference with a zero radius.
The integrals along the circumferences are going to the zero. The integrals along
the straight lines give:
$$ \int [ \ln^{j+1} y - ( \ln y + 2\pi i )^{j+1}]
\frac{
\sum_{i=0}^{n-1} \frac{y^i H^{i+1}}{(i+1)!}
}{
\sum_{i=0}^{n} \frac{y^i}{i!}
\sum_{i=0}^{n} \frac{y^i H^i}{i!}
}
dy $$
which can be reduced to $I_j$ and $\sum_{k<j} e_k I_k$ with
some known coefficients.
On the other hand this integral can be reduced to
$$
\sum res(\ln^{j+1} y
\frac{
\sum_{i=0}^{n-1} \frac{y^i H^{i+1}}{(i+1)!}
}{
\sum_{i=0}^{n} \frac{y^i}{i!}
\sum_{i=0}^{n} \frac{y^i H^i}{i!}
}
)$$
inside $\Omega$. As the result we have the recurrent procedure which allows
us to calculate all $I_j$.

Now let us see how the corrections due to the cut-off can be introduced  in
the value of $g(z,w)$. In the first approximation we have:
$$
g_{1}(z,w) = \frac{6 f_*}{c^4} \exp(\lambda w) (\exp(cz) - \exp(-ca))
$$
The value of $G_1$ in the approximation $\eta_{tot} = const$ is given by
$$
G_1 = f_{\zeta\ *}\frac{ 6 \eta_{tot}}{ c^4} \frac{1}{A \lambda}
(\exp(cz) - \exp(-ca))
$$
Note that it can be calculated for the finite band of
 the spectrum of the activities. The
calculation under the polinomial approximation for $\eta_{tot}$ can be
done in the same manner.

Instead of the pre-exponential factor one can put  some other
function. The separation of this expression into two factors depended on
$w$ and on $z$ is obvious.

For $\theta_2$ we get:
$$\theta_2(z,w) =
\exp(- f_{\zeta \ *} n_{\infty} \exp(\lambda w)
\int_{-\infty}^{z} \exp(cx - B(\exp(cx) - \exp(-ca)) ) dx )
$$
and after the simple calculation:
$$
\theta_2(z,w) =
\exp(-f_{\zeta \ *} n_{\infty} \exp(\lambda w) \exp(B\exp(-ca))
\frac{1}{cB}(1-\exp(-B \exp(cz))))
$$

The final value is the following:
  $$
\theta_2(z,\infty) =
\exp(-f_{\zeta \ *} n_{\infty} \exp(\lambda w) \exp(B\exp(-ca))
\frac{1}{cB})
$$
It differs from (\ref{*}) only by the remormalisation of the amplitude
thanks to $\exp(B \exp(-ca))$.

The value of $N^{tot}$
is calculated in the same manner.

Note that one can change $-\infty$ to $-a$ and repeat all calculations
in the same manner.

The analytical estimates show that the second iteration gives rather a good approximation
for the process. The reason is quite similar to \cite{Novosib}.
Really, one can simply integrate all estimates from \cite{Novosib} to
justify the validity of the obtained expressions.

Another important remark must be given. One can see that all obtained
values have the separation  of the expressions (or arguments of exponents) in
two terms:
the first one depended on $w$ and
the second one  depended on $z$.
This reduction can not be observed in
all high iterations. It is responsible
for the absence of the cross influence of exhaustion of the heterogeneous
centers with the different activity in the first iterations.
The similar property is absent in the
situation of decay \cite{aero1}, \cite{aero2}. Here such a cross influence
can be neglected because
the
equation on
the parameters of condensation
balances the time of formation with
the characteristic intensity of formation
and leads to the fact that the characteristic
width of the size spectrum is always
approximately equal to $\frac{1}{c}$.
The probability to form the droplet on the center is always determined
only by the supersaturation and by the value of activity
(and not by the total number of centers of
the given
sort).

\section{Universal solution}

Now the problem of the construction of the universal solution will
be considered. The idea of the universal solution
\cite{13} lies in the fact that after formation of the spectrum of
the droplets
the further evolution of the process depends only from the first three
(and zero) momentums of the distribution function. So if one can choose some
variables in which the solution
(the distribution function) is the universal function (undepended on
the parameters of the problem)
then the expressions for the momentums have rather a simple analytical structure
combining some parameters of the problem with    the
universal constants.

For the process of the homogeneous condensation
the universal solution was obtained in \cite{13}. For
the process of the heterogeneous
condensation
the universal solution is absent, but some pseudo universal solution can be
considered  as the base for
the further iterations \cite{PhysRev}.

We shall rewrite
the system of the  condensation equations in the terms of $\zeta$ and $\Phi$
$$
\Phi_* + \frac{\Phi_*}{\Gamma} c z = \zeta + G
$$
$$
G = \int dw g(w)
$$
$$
g(w) = f_* \int_{-\infty}^{z} (z-x)^3 \exp(\frac{\Gamma}{\Phi_*}
(\zeta - \Phi_*)) \theta(w,x) dx \exp(\lambda w)
$$
$$
\theta(w) = \exp(-f_{\zeta\ *} n_{\infty} \exp(\lambda w)
\int_{-\infty}^{z} \exp(\frac{\Gamma}{\Phi_*} (\zeta - \Phi_*)) dx)
$$
Introduce the function
$$
\delta = \frac{\Gamma}{\Phi_*} (\zeta - \Phi_*)
$$
Then after the substitutions
$$
\lambda w \rightarrow w
$$
$$
px \rightarrow x
$$
$$
\frac{\Gamma}{\Phi_*} G \rightarrow G
$$
$$
\frac{\Gamma}{\Phi_* \lambda } g \rightarrow g
$$
one can get
$$
\frac{ c}{p} z = \delta + G
$$
$$
G = \int dw g(w)
$$
$$
g(w) =
\frac{f_* \Gamma}{\lambda \Phi_* p^4}
 \int_{-\infty}^{z}
 (z-x)^3 \exp(\delta) \theta(w,x) dx \exp(w)
$$
$$
\theta(w) = \exp(-f_{\zeta\ *} \frac{n_{\infty}}{p} \exp(w)
\int_{-\infty}^{z} \exp(\delta) dx)
$$
Let us choose $p$ as
$$
\frac{f_* \Gamma}{\lambda \Phi_* p^4} = 1
$$
and now there are no
parameters in the expression for $g$.
If the moment of $t_*$ is chosen as the moment when the supersaturation
attains                                                 the
maximum, then
$$
\frac{c}{p} =
\int dw 3  \int_{-\infty}^{z}
 (z-x)^2 \exp(\delta) \theta(w,x) dx \exp(w)
$$
Then
$$
f_{\zeta \ *} \frac{n_{\infty}}{p} =
\frac{
  f_{*}^{3/4} n_{\infty}   \lambda^{1/4} \Phi_*^{1/4}
}{
\eta_{tot}
\Gamma^{1/4}
}
$$

An ordinary and natural condition to establish the zero point of activity
can be written as the following one:
$$
\theta(w=0,z=\infty)
= \frac{1}{2}
$$
which gives
$$
f_{*} \frac{n_{\infty}}{  p } =
\frac{\ln 2}{\int_{-\infty}^{\infty} \exp(\delta(x)) dx}
$$
and there are no parameters in the last equation.

As the result there are no parameters in the system of
the condensation equations and the
solution
has the universal form.
It can be analytically proved that it is a unique solution. All consequences now
coincides with analogous conclusions from
\cite{13}

Note that
there is a property of a very smooth dependence of the number of droplets on the
$f_*$. So we can get the
equation on the parameters  of the process
in some rough approximation
(see the iteration procedure) and then use the universal law.

The last model has the evident disadvantage. The equation for $G$  demonstrates
no convergence. Really, we can substitute the expression for $g$ into
the expression for $G$  and fulfil the integration over $w$.  Then we
come to
$$
G =                                                          A_{00}
\int_{-\infty}^z (z-x)^3
\frac{\exp(\delta (x)) }{\int_{-\infty}^x \exp(\delta(x')) dx' } dx
$$
where
$$
A_{00} =
\frac{f_* \Gamma}{\lambda \Phi_* p^3 n_{\infty} f_{\zeta\ *}}
$$
The evident necessary asympote $G \rightarrow 0$ when $z \rightarrow
- \infty$ leads to  $ \delta (z) \rightarrow \frac{c}{p} z $ when $z \rightarrow
 - \infty$. Then the exponents can be cancelled
and the subintegral expression has the asymptote $(z-x)^3$
which doesn't ensure the convergence.  Then we need some regularizations.
The most evident one is to  notice that after the process of any appearing
of the droplets the constant value of $\eta$ transforms into
$$ \eta_{new} = \eta_{old} \exp(-Q \exp(w)) $$
where $Q$ is some constant.

We shall use this value as the initial one and after the analogous transformations
we can come to
$$
G=
\frac{f_* \Gamma}
{\lambda \Phi_* p^4}
\int_{-\infty}^z
(z-x)^3
\frac{\exp(\delta (x)}
{Q+\frac{f_* n_{\infty}}{p} \int_{-\infty}{z} \exp( \delta (x'))dx'} dx
$$
The coefficient $          \frac{f_* n_{\infty}}{p} $
can be made  equal to some constant (may be 1) due to the choice of
the scale of activities.  This choice  states some concrete value of $Q$
(which has the same dependence on the choice as $\frac{f_* n_{\infty}}{p}$
has). The limit $Q \rightarrow 0$ corresponds to the already observed situation.
Now we have more general equation taking into account the power of the
previous deformation of the activity spectrum. Evidently the solution
depends on the parameter $Q$ and isn`t universal. We see the further inclusion
of the distribution into some more general set of the solutions for
the various powers
of the deformations of the spectrums by the previous nucleation.

Due to the problems of convergence for $Q=0$ we must describe the form
of the spectrum explicitly. We are going to  show that the size spectrum
has the  universal character.

When $Q$ is going to zero the region where $Q$ is unessential starts at
$z_l$ corresponding to
$$  \int_{-\infty}^{z_l} \exp(\delta(x)) dx
\sim (p/c) \exp( cz_l / p) \approx Q
$$
and becomes larger. The asymptote $(z-x)^3$ of the subintegral
expression is attained from $z_l$ till some  $z$ with no respect
to $Q$. So, the asymptotic region becomes larger. Certainly, we can integrate
the asymptote explicitly and come to $(z-z_l)^4/4$. Only due to the
linear size this asymptote provides
the main quantity of the substance in the droplets. As a result we have
the following expression for the behavior of $f_{\zeta  }$
$$
f_{\zeta  } \sim \exp( z - \alpha ( z-z_l )^4 )
$$
where
the scale of $z$ is chosen to put $c/p = 1$ and $ \alpha$ is some constant.
The value of $\alpha$ can be determined from the requirement that
$ max \ f_{\zeta \ *} $ (the maximum of the supersaturation)
is attained at $z=0$. Then
$$
\alpha = - z_l^{-3} / 4
$$
The expression for $f_{\zeta \ *}$ can be presented in the following form
$$
f_{\zeta \ * } \sim \exp( -\alpha z_l^4)
\exp(-(z/\Delta)^2) \exp(-4\alpha z^3 | z_l |) \exp(-\alpha z^4)
$$
where
$\Delta$ is the characteristic halfwidth
$$
\Delta = (2 |z_l| /3 )^{1/2}
$$
One can see that at the characteristic scale $z \approx \Delta$ the arguments
of the third and the forth exponents have the characteristic  values $z_l^{-1/2}$
and $z_l^{-1}$ respectively.
Thus, these terms can be neglected and the form  of $f_{\zeta \ *}$ is
the gaussian                                            one:
$$
f_{\zeta \ * } \sim
\exp(-(z/\Delta)^2)
$$
and it has no parameters after the evident rescaling\footnote{This rescaling
excludes the unphysical parameter $z_l$.}
$z \rightarrow z/\Delta$. So, we see that
the universal gaussian form of $\exp(\delta)$ ensures the universal expression
for $\theta$
$$
\theta = \exp( - \exp(w) \frac{ \ln 2}{\sqrt{\pi}} \int_{-\infty}^z \exp(-x^2)
dx)
$$
This universal form will be attained earlier then $Q=0$ (where the property
of convergence disappears) and has the true physical sense.

Figure 1 shows the
forms of the size spectrums in some
different situations presented in the normalized
coordinates. The curves "a", "b", "c", "d" are the solutions of
the following equation
$$
z = A_1 \int_{-\infty}^z (z-x)^3
\frac{\exp(\delta(x))}{Q + \int_{-\infty}^x \exp(\delta(x')) dx'}
dx
$$
for the different values of  $Q$ (
"a" : $Q=10$
"b" : $Q=1$
"c" : $Q=0.001$
"d" : $Q=0.0001$
). The curve "a" practically coincides with "b",
the curve "c" practically coincides with "d" for all $x$.
The curve "e" is  the spectrum of the gaussian type
$\exp(-x^2)$. The curve "f" is the universal form of the first
iteration $\exp(x-\exp(x))$ after the renormalization
in order to have the same position of
the maximum of the curve and to conserve the integral  over
the spectrum. This curve practically coincides with "a", "b" but only for the
negative values of the argument.

One can prove that for all possible values of $Q$ the spectrum lies between
the first iteration and the Gaussian spectrum. It can be seen that
the difference between the first iteration and the Gaussian curve  is
rather small (less than $0.2$)
and lies in the frames of the accuracy of the modern experiment\footnote{This
ensures the weak sensitivity to the choice of the parameter $a$. It is
necessary to the self-consistency of the presented theory.}.
 Note that the value of $A_1$ must be  chosen to ensure the position
of the maximum of the supersaturation near $max\ \delta = 0$. For
$Q \rightarrow 0$ the quantity of the substance in the droplets isn't small
due to the known power asymptote and one have to remove the choice of $t_*$
from the previous
condition that the $max \ \delta$ is attained at $z=0$ to the new condition
$max \ \delta = 0$.
In the concrete calculations one needn't to fulfil this condition precisely
 but only in a very approximate way as far as the form of the spectrum
doesn't change essentially for the different choice of the $max \ \delta$
near zero.
This property can be shown analytically and is illustrated by  Figure 2.

Figure 2 demonstrates the weak dependence of the form of spectrum on the
choice of the base point for the linearizations, i.e. the choice of $t_*$.
This property is important for the justification of the conception of
the universal spectrum and can be proved analytically.
This property is also necessary
 for the simplification of the numerical
calculations. For $Q=10$ the two values of $A_1$ have been
chosen. This corresponds to the different manners of the choice of $t_*$
or
the choice of $max\ \delta$. Two curves for $\delta$ for different $A_1$
are shown in the part "a" of this illustration. The upper curve corresponds
to $A_1 = 0.5$, the lower curve - to $A_1 = 1$. In the part "b" the size
spectrums for these situations are drawn in the normalized coordinates.
They coincide and one can see only one curve drawn in the part "b".

Note that the limit $Q \rightarrow 0$ is analogous to the situation of
the "wide spectrum" considered in \cite{aero2}. Here we have some
additional difficulties. These difficulties appear from the asymptote
$(z-x)^3$ for the subintegral expression. As a result the
relative quantity of
the substance in the droplets is greater than $\Gamma^{-1}$, the spectrum
is going to be formed  during the period with duration essentially longer
than $c^{-1}\tau / \zeta$, the quasistationary approximation isn't valid
during all the period of the droplets formation. So, one has to reconsider
the approximations which form the base of the concrete constructions.

The simplification goes from the asymptote $(z-x)^3$. Really, this asymptote
doesn't depend on the concrete form of the free energy and the derivatives
of the free energy. So, the quasistationarity
and  exponential approximation (\ref{expap}) aren't so essential (they
are going to  fail
namely at this asymptote).
The big quantity of the substance in the droplets require to reconsider
the condition of the choice of $t_*$ as it is done above.
The essential duration of the period of the droplets
formation may lead to  some new approximations and, thus, change the form
of the asymptote $(z-x)^3$. But the consideration of the process remains
quite analogous to the already fulfilled one. One of such examples is given
in the next section.

The opposite situation corresponds  to the validity of the iteration procedure
with the ideal supersaturation  taken as the external
one\footnote{Certainly,
with the corresponding choice
  of the renormalization of $\theta_{init}$
and the choice of the "external" supersaturation the iteration procedure
remains valid in all cases with the linear "external" supersaturation.}. Note
that the final results for $\theta$ and for the spectrum are obtained
(in the second approximation) on the base of the supersaturation calculated
without the appropriate account of the exhaustion of the centers, i.e.
in  the pseudohomogeneous situation. The same situation occurs when we
study the condensation on the separate  sort
of the heterogeneous centers \cite{Novosib},
\cite{PhysRev}. This analogy leads to the  three important consequences.
To  establish the first one note that the "pseudohomogeneous" base for
the final results evidently has some universiality. In \cite{PhysRev}
the universiality can be broken by the change of the regime  of the consumption
of the vapor by the droplets. The result of \cite{Novosib} is more approximate,
but it is based on the form of the first iteration which can not be depended
on the regime of the vapor consumption. So, the universiality  here is
more general. The  analogy of the  forms of the "pseudohomogeneous" spectrums
allows to give the definition of the relative  activity also to the centers in the
the process of condensation on the centers of the separate sort. The point
$w=0$ corresponds to $\theta_{final} = 1/2$. The value of $w$ can be reconstructed
from $\theta_{final}$ as
$$ w = \ln(  \frac{ \ln \theta_{final}    }{  \ln(1/2)    }  )  $$
The invariant $\ln \theta / \exp(w)$ can be also observed for an arbitrary
$z$. Hence, the condensation occurs in a
hierarchical manner\footnote{Certainly,
the shift and the zero roint $w=0$ depend on the parameters of the process
including $\eta_{tot}$.}. Then one can  consider the distribution $f(x,w)$
over the two variables $x$ and $w$ obtained in the process of condensation
on the heterogeneous centers with the spectrum of activities as the formal
generalization of the distribution in the  process of condensation on
the  heterogeneous centers of the separate sort. Now
the distribution has the  universal functions as the
base. This base doesn't depend on the
 process of the exhaustion of the heterogeneous centers. The dependence
on $w$ is rather explicit
$$
 f(x,w) \sim \exp(\delta(x)) \exp(-w \int_{-\infty}^{x} \exp(\delta(x'))
dx'
\frac{\ln 2}{\int_{-\infty}^{\infty} exp(\delta (x') ) dx' } )
$$
The factor $\exp(\delta)$ is the universal base for this distribution.
That's why we prefer to present it in Figure 1 and to call it simply the
spectrum.  The last expression is precisely valid  for $\delta$ calculated on
the base of the precise solution. As for approximations for $\delta$ one
can take $\delta$ from the pseudohomogeneous situation in two manners
(from \cite{Novosib} and from \cite{PhysRev}). The first one corresponds
to the first  iteration and doesn't  depend on the  regime of the vapor
consumption. The second  one is more precise but depends on the regime
of the vapor consumption. Note that  the  last expression contains an
unessential approximation because it is based on the pseudohomogeneous
situation. The negligible character of correction can be seen simple from
the result of the iteration procedure in the situation with one sort of
the heterogeneous centers. Hence, one can state the natural inclusion
of the specific condensation on a sort of the heterogeneous centers into
the practically universal distribution appeared from the process of the
condensation on the heterogeneous centers with the spectrum of activities.

The second consequence  is connected with the fact that now we know the
ansatz necessary for the realization of the first opportunity (alternative
to the regularization of $\theta$). One can use  the form of $\delta$
obtained in the pseudohomogeneous situation in the first iteration or from
the universal homogeneous precise solution.

The third  consequence can be noticed in frames of the modified method of
the steepens descent \cite{aero3}. Really, the form of the external supersaturation
can be essentially nonlinear\footnote{In a very specific global form of the
activity spectrum $\eta_{tot}$. One can prove that in any case one can
restrict the expression for the external supersaturation during the
period of the intensive formation of the droplets on the centers with the intermediate activity
by the constant, linear and square terms.}. Then
we have to use the modified method
of the steepens descent. The generalization is evident. Due to the negligible
effect of the account of the exhaustion of the heterogeneous centers in
all steps except the final formulas which can be seen  from the iteration method
(in the nonlinear case all analogous estimates can be given) we can simply
integrate (summarize) the amplitudes of the distribution\footnote{The
characteristic values of the distribution at $t_*$.} over the activity,
then get the result for the psudohomogeneous situation and then apply
the final formulas for the number of the heterogeneous centers on the
base of the supersaturation.

One can analytically show for $Q \geq 1$ that the differences
in the forms of the spectrums between the   real solution
and the first iteration, between the real solution and the pseudohomogeneous
universal precise solution
 decrease when $Q$ increases. Also one can prove that    for $Q
\geq 1$
the difference between the external supersaturation and the real supersaturation
at some arbitrary moment of time decreases\footnote{And also at $t_*$.}
when
$Q$ increases.
  So, the results of Figure 1 show
that the required limit corresponding to the practically universal spectrum
is attained already when $Q \geq 1$.

One can analytically show that the form of the spectrum lies between the
gaussian iteration and the first iteration (or
between the gaussian form and the universal precise pseudohomogeneous
solution).  We can see from Figure 1 that the difference in the form of
the spectrum is rather small for all situations. This can lead us toward
the general approximate universiality. So, the analogy with the ideas
of the universiality in the previous considerations is now stated.

Certainly, in the case of the condensation on the  separate sort of the
heterogeneous centers the spectrum (as $\theta \exp(\delta)$) isn't universal.
But the difference between the forms of the spectrum isn't so important.
One can show analytically  that the form\footnote{After renormalization
and as the function of a rescaled variable.} of the spectrum changes
continiously from the case of the relatively small number of the heterogeneous
centers up to the  pseudohomogeneous case and lies between these two limit
cases.  Really,  when the relative number of the heterogeneous centers
is small the first  iteration $\exp(x - \exp(x))$ gives the precise expression
for the spectrum. In the pseudohomogeneous case the same first iteration
is the base  for the result which lies prctically near the precise solution
(this is the base point for the effectiveness of the iteration method
\cite{Novosib}). So, the form of the spectrum doesn't essentially variate.
It can be also treated as as some universal form. Hence, this case also
allows the universal description. Namely this is the base for the applicability
of the iteration procedures from \cite{Novosib}, \cite{PhysRev}.

\section{Advantages of the model}

The results of the iteration procedure can give not only the description
of the process of the heterogeneous
condensation but also the information about the form
of the activity spectrum. From the first point of view this question appears
to be an external one to the process of condensation.

Experimental results
\cite{Twomey}
show that the spectrums of the activities
are rather smooth and have the form
$$ \varphi (v) \sim v^{-(1+s)} $$
where $s$ is some small positive parameter. Here $v$ has the sense of
the activity introduced in a slightly another manner.
Namely, the free energy can be approximately written
as the function of $v$ in the leading term   as
$$
F =  - \ln(\zeta+1) \rho^3 + const \rho^2 + const v \ln \nu + const
$$
which leads to the approximate applicability of  the linearization of
$\Delta F = F_c - G = F(\rho_c) - F(\rho_e) $
as function of $w$ together with the identification  of $v$ as $w-w_0$ with
some characteristic value of the parameter $w_0$. Then $\eta_{tot}$ can be
presented as
\begin{equation} \label{h**}
\eta_{tot} (w) = \eta_{tot} (w_0) (w-w_0)^{-(1+s)}
\end{equation}
The positive value of $s$ ensures  the convergence of the total number
of the heterogeneous centers\footnote{The infinite tail must be cut off
at the centers which must be already exhausted in the previous processes.}
$$
\eta^{tot} = \int_{w}^{\infty} \eta_{tot} (w) dw
$$
Define the class of the "long tail spectrums" (LTS) as the spectrums with extremely
long tails in the active  region. This spectrum shows an example of such
a spectrum. Here we shall develop the theory for condensation on such
spectrums.

To present the most simple variant we shall assume that for $f_{\zeta_{*}}$
an approximation of the known form is adopted
\begin{equation} \label{h***}
f_{\zeta_{*}} = f_{\zeta_{**}} \exp(\frac{\Gamma}{\zeta_{**}} (\zeta_*
- \zeta_{**}))
\end{equation}
with some constant value of the known parameter $\Gamma$. Here
$\zeta_*$ can be treated as a value corresponding to the moment when some
sort of the heterogeneous centers is exhausted and $\zeta_{**}$ is the
base for approximations. Require
that
$$
\theta (z=0, w) = 1/2
$$
which leads to
$$
\frac{\Gamma}{\zeta_{**}} (\zeta_* - \zeta_{**}) + \lambda w =
- \ln[\frac{f_{\zeta_{**}}n_{\infty}}{c \ln 2}]
$$
The last approximation is valid when the relative variation of $\zeta$
is small. In such a region we can put $c$ in the r.h.s. of the last equation
to some  constant value and get for the size spectrum
$$
\mid f(z) \mid = \mid \eta_{tot}(w\mid_{w=\frac{const - cz}{\lambda}})
\mid \frac{c}{\lambda}
$$
where the dependence $\ln   c$ is assumed to be unessential.
Then one can  see that in the case of (\ref{h**}) the convergence of the
integral for $G$ can be attained only if $s>3$ (note that $c \rightarrow
const \sim  \frac{d\Phi}{dx}$ when $w \rightarrow \infty$). This result
 shows that LTS can not be effectively  spread. More rigorously speaking,
the contradiction can be overcome by noticing that this effect is only
due to  approximation (\ref{h***}). Certainly, $G$ is restricted by
the value $\eta_{tot} z_{max}^3$ where $z_{max}$ is the coordinate of
the  droplets which are imaginary
formed when the supersaturation attains some  slightly positive value.
Nevertheless in such a situation all global features of the free energy appear
which doesn't allow to get any effective method to all types of the heterogeneous
centers.
The leading idea will be that some of heterogeneous centers had
been exhausted in the previous processes of condensation occurred earlier.
Then the balance equation will be  the following
$$
\frac{d\Phi}{dx} = c \frac{\Phi}{\Gamma} + 3 \int_{-d}^{z} (z-x)^2 f(x)
dx
$$
where $d$ is some boundary parameter of size spectrum initiated by boundary
of activity spectrum. This equation can be solved. In the case of $\eta_{tot}
\sim const$ we have
\begin{equation} \label{h++}
\frac{d\Phi}{dx} = c  \frac{\zeta_{**}}{\Gamma} + 3
\int_{-d}^{z} (z-x)^2 c(x) dx \frac{const}{\lambda}
\end{equation}
Note that here $c\frac{\zeta_{**}}{\Gamma}$ is the derivative of the real
(not the ideal) supersaturation on $x$.  Having differentiated this equation
three times\footnote{Now we
are going to get the equation for the unknown function $c(x)$ so we have
to keep the derivatives of $c(x)$} we
get the ordinary linear differential equation with some constant
coefficients,
which has the known solution.

Now a new principle of the self-consistency of the activity spectrum will
be elaborated. We have noticed that active centers have been  exhausted
in the previous processes of condensation. But the description of the
previous processes of condensation is quite analogous to the description
of the given process. As the result we have for spectrum of
the heterogeneous centers
$$
\theta = \exp( - const \exp(\lambda w)
\int_{-\infty}^{z}
\exp(-\Gamma \frac{\zeta
- \Phi_*}{\Phi_*} ) dx )
$$
with the analytical structure
\begin{equation} \label{h+++}
\theta ( z \rightarrow \infty )  = \exp(-const \exp(\lambda w))
\end{equation}
But this structure is already known - it is the structure of the final and
the start results for $\theta$ in the already investigated process. So, our
start form was absolutely correct\footnote{One can repeat all calculations
with the new value of parameters initiated by the previous process of
consumption of the heterogeneous centers.}. Also it is very important that the
structures before and after the process coincide in their analytical form.
This statement will be called as the principle of the self-consistency of
the spectrum of the activities.

Hence, our initial conditions satisfies this principle. The process of
condensation conserves the analytical form of the  spectrum
of the activities  and the result of the process  can
be regarded as some shift in the scale of activities.
Only some  parameters  are
changed.
We needn't to repeat the calculations as far as they are absolutely the same
ones.

However, in (\ref{h++}) when $d \gg 1$ we needn't
to know the details of the cut-off of
the spectrum  and can use the cut-off by the Heavisaid's function.

The last point of discussion is the possibility of
the linearization of the external
supersaturation. One can analytically prove the following statement:
\begin{itemize}
\item
In the  description of condensation  on the centers of the intermediate
activity the two manners of description cover all possible
situations\footnote{The natural requirement for the activity
spectrum is that the activity
spectrum must be  a smooth function with the cut-off (or without
the cut-off) like
(\ref{h+++}).}. These
manners of description are the following:

1). The supersaturation is  absolutely determined by the consumption of
the vapor
by the droplets  formed on the centers with high activity.

2). The linearization of the external supersaturation is possible during the
period  of  formation  of  the  droplets  on   the   centers   with
the intermediate activity.

\end{itemize}

The first manner of description is obviously trivial. The second is completely
described here. Note that  in the case when the spectrum has some pseudo
singularities
the last statement isn't valid and the special description is necessary.
This description can be attained by the combination of the methods presented
in \cite{aero3} with this theory.

\section{Concluding remarks}

Now the theory  of aerosol formation is completed. As a base for some
concrete results the classical theory of nucleation was chosen. Meanwhile
the validity of this theory remains the matter of discussion\footnote{Note
that formally we can put the discrepancy between the predictions of the
classical theory and the real rate of nucleation into the values of the
microscopic corrections.}. Note that this theory is necessary only to
calculate the amplitude of the distribution. In the situation of the dynamic
conditions the manner of the choice of $t_*$  leads to the algebraic equation
of the following form:
$$ f_* = Smooth \ known \ function (\Phi_*; explicit \ external \ parameters) $$
So, the microscopic  corrections to the free energy aren't important  for
the process of condensation under the dynamic conditions. In the situation
of the metastable phase decay the amplitude value of the spectrum is rather
artificial. It is given explicitly by  the initial supersaturation and
one can  not exclude the sharp dependence on the microscopic corrections
to the free energy. Here any other  concrete formula can be used instead
of the  expression from the classical theory for the nucleation rate in
the stationary approximation. Note that  the form of the spectrum doesn't
essentially depends on the value of the amplitude. That's
why we don't pay  any serious attention to the solution of the algebraic
 equations on the parametrs of the process and to the calculation of the
concrete numerical results. We have concentrated our  attention on the
universal depencies which form the base for the common knowledge in the
field of the first order phase transitions.  Namely the universiality
is the main result of our investigation. Note that the universiality also
takes place in the process of the condensation under the dynamic conditions.

The theory presented in these four publications can be easily reformulated
when we assume that the rate of appearance of the critical embryos is
proportional to
$$ F_1 \exp(F_2) $$
where $F_1$ and $F_2$ are the smooth functions  of the supersaturation
and $F_2$ has the  big absolute  values. Evidently all reasonable recipes
for the rate of nucleation satisfyes the last ansatz. We kept the classical
stationary flow only due to the tradition. The range of applicability of the
presented theory  is wider than the case of condensation. The possibility
 of the linearization of the ideal power of metastability during the period
of the intensive formation of the super-critical embryos seems to be quite
natural. The sharp increasing of the intensity of the metastable phase
consumption by the  separate super-critical embryo is also a
natural requirement. This theory is based only upon these assumptions.
All positions used in the construction of the mathematical  model  (exept
the absence of coalescence) can be missed\footnote{The thermal effects
can be taken into account in the manner analogous to \cite{Novosib}.}
 by the obvious generalization
conserving the mathematical structure. The requirement of the homogeneous
character of the external  action and of the nucleus distribution can be
attained by the consideration of an arbitrary
 hydrodynamic element. As the result
one can conclude that the general theory for the kinetics of formation
of the dispersed embryos of a new phase on the dispersed unpurities in
a metastable phase is constructed.

\pagebreak

\begin{picture}(350,300)
\put(80,250){f=a=b,c=d,e}
\put(250,250){f,a=b,c=d,e}
\put(50,45){.}
\put(52,45){.}
\put(53,45){.}
\put(54,45){.}
\put(55,46){.}
\put(57,46){.}
\put(58,46){.}
\put(59,46){.}
\put(60,46){.}
\put(61,47){.}
\put(63,47){.}
\put(64,47){.}
\put(65,48){.}
\put(66,48){.}
\put(68,48){.}
\put(69,49){.}
\put(70,49){.}
\put(71,49){.}
\put(73,50){.}
\put(74,50){.}
\put(75,50){.}
\put(76,51){.}
\put(78,51){.}
\put(79,52){.}
\put(80,52){.}
\put(81,53){.}
\put(83,53){.}
\put(84,54){.}
\put(85,54){.}
\put(86,55){.}
\put(88,55){.}
\put(89,56){.}
\put(90,57){.}
\put(91,57){.}
\put(93,58){.}
\put(94,59){.}
\put(95,59){.}
\put(96,60){.}
\put(98,61){.}
\put(99,62){.}
\put(100,63){.}
\put(101,63){.}
\put(102,64){.}
\put(104,65){.}
\put(105,66){.}
\put(106,67){.}
\put(107,68){.}
\put(109,69){.}
\put(110,70){.}
\put(111,72){.}
\put(112,73){.}
\put(114,74){.}
\put(115,75){.}
\put(116,77){.}
\put(117,78){.}
\put(119,79){.}
\put(120,81){.}
\put(121,82){.}
\put(122,84){.}
\put(124,86){.}
\put(125,87){.}
\put(126,89){.}
\put(127,91){.}
\put(129,93){.}
\put(130,95){.}
\put(131,97){.}
\put(132,99){.}
\put(134,101){.}
\put(135,103){.}
\put(136,105){.}
\put(137,107){.}
\put(139,109){.}
\put(140,112){.}
\put(141,114){.}
\put(142,117){.}
\put(143,119){.}
\put(145,122){.}
\put(146,125){.}
\put(147,127){.}
\put(148,130){.}
\put(150,133){.}
\put(151,136){.}
\put(152,139){.}
\put(153,142){.}
\put(155,145){.}
\put(156,148){.}
\put(157,151){.}
\put(158,154){.}
\put(160,158){.}
\put(161,161){.}
\put(162,164){.}
\put(163,168){.}
\put(165,171){.}
\put(166,174){.}
\put(167,178){.}
\put(168,181){.}
\put(170,185){.}
\put(171,188){.}
\put(172,191){.}
\put(173,195){.}
\put(175,198){.}
\put(176,201){.}
\put(177,204){.}
\put(178,208){.}
\put(180,211){.}
\put(181,214){.}
\put(182,217){.}
\put(183,219){.}
\put(184,222){.}
\put(186,224){.}
\put(187,227){.}
\put(188,229){.}
\put(189,231){.}
\put(191,233){.}
\put(192,235){.}
\put(193,236){.}
\put(194,237){.}
\put(196,238){.}
\put(197,239){.}
\put(198,240){.}
\put(199,240){.}
\put(201,240){.}
\put(202,240){.}
\put(203,239){.}
\put(204,238){.}
\put(206,237){.}
\put(207,236){.}
\put(208,234){.}
\put(209,232){.}
\put(211,229){.}
\put(212,227){.}
\put(213,224){.}
\put(214,220){.}
\put(216,217){.}
\put(217,213){.}
\put(218,209){.}
\put(219,205){.}
\put(220,200){.}
\put(222,196){.}
\put(223,191){.}
\put(224,185){.}
\put(225,180){.}
\put(227,175){.}
\put(228,169){.}
\put(229,164){.}
\put(230,158){.}
\put(232,152){.}
\put(233,147){.}
\put(234,141){.}
\put(235,135){.}
\put(237,130){.}
\put(238,124){.}
\put(239,119){.}
\put(240,114){.}
\put(242,109){.}
\put(243,104){.}
\put(244,99){.}
\put(245,94){.}
\put(247,90){.}
\put(248,86){.}
\put(249,82){.}
\put(250,78){.}
\put(252,74){.}
\put(253,71){.}
\put(254,68){.}
\put(255,65){.}
\put(257,62){.}
\put(258,60){.}
\put(259,57){.}
\put(260,55){.}
\put(261,54){.}
\put(263,52){.}
\put(264,50){.}
\put(265,49){.}
\put(266,48){.}
\put(268,47){.}
\put(269,46){.}
\put(270,45){.}
\put(271,44){.}
\put(273,43){.}
\put(274,43){.}
\put(275,42){.}
\put(276,42){.}
\put(278,42){.}
\put(279,41){.}
\put(280,41){.}
\put(281,41){.}
\put(283,41){.}
\put(284,41){.}
\put(285,40){.}
\put(286,40){.}
\put(288,40){.}
\put(289,40){.}
\put(290,40){.}
\put(291,40){.}
\put(293,40){.}
\put(294,40){.}
\put(295,40){.}
\put(296,40){.}
\put(298,40){.}
\put(299,40){.}
\put(300,40){.}
\put(301,40){.}
\put(302,40){.}
\put(304,40){.}
\put(305,40){.}
\put(306,40){.}
\put(307,40){.}
\put(309,40){.}
\put(310,40){.}
\put(311,40){.}
\put(312,40){.}
\put(314,40){.}
\put(315,40){.}
\put(316,40){.}
\put(317,40){.}
\put(319,40){.}
\put(320,40){.}
\put(321,40){.}
\put(322,40){.}
\put(324,40){.}
\put(325,40){.}
\put(326,40){.}
\put(327,40){.}
\put(329,40){.}
\put(330,40){.}
\put(331,40){.}
\put(332,40){.}
\put(334,40){.}
\put(335,40){.}
\put(336,40){.}
\put(337,40){.}
\put(339,40){.}
\put(340,40){.}
\put(341,40){.}
\put(342,40){.}
\put(343,40){.}
\put(345,40){.}
\put(346,40){.}
\put(347,40){.}
\put(348,40){.}
\put(350,40){.}
\put(50,44){.}
\put(51,44){.}
\put(53,44){.}
\put(54,45){.}
\put(55,45){.}
\put(56,45){.}
\put(57,45){.}
\put(59,45){.}
\put(60,46){.}
\put(61,46){.}
\put(62,46){.}
\put(63,46){.}
\put(64,47){.}
\put(66,47){.}
\put(67,47){.}
\put(68,47){.}
\put(69,48){.}
\put(70,48){.}
\put(72,48){.}
\put(73,49){.}
\put(74,49){.}
\put(75,49){.}
\put(76,50){.}
\put(78,50){.}
\put(79,51){.}
\put(80,51){.}
\put(81,51){.}
\put(82,52){.}
\put(84,52){.}
\put(85,53){.}
\put(86,53){.}
\put(87,54){.}
\put(88,54){.}
\put(90,55){.}
\put(91,56){.}
\put(92,56){.}
\put(93,57){.}
\put(94,58){.}
\put(96,58){.}
\put(97,59){.}
\put(98,60){.}
\put(99,60){.}
\put(100,61){.}
\put(101,62){.}
\put(103,63){.}
\put(104,64){.}
\put(105,65){.}
\put(106,66){.}
\put(107,67){.}
\put(109,68){.}
\put(110,69){.}
\put(111,70){.}
\put(112,71){.}
\put(113,72){.}
\put(115,73){.}
\put(116,75){.}
\put(117,76){.}
\put(118,77){.}
\put(119,79){.}
\put(121,80){.}
\put(122,82){.}
\put(123,83){.}
\put(124,85){.}
\put(125,86){.}
\put(127,88){.}
\put(128,90){.}
\put(129,91){.}
\put(130,93){.}
\put(131,95){.}
\put(133,97){.}
\put(134,99){.}
\put(135,101){.}
\put(136,103){.}
\put(137,106){.}
\put(139,108){.}
\put(140,110){.}
\put(141,112){.}
\put(142,115){.}
\put(143,117){.}
\put(144,120){.}
\put(146,122){.}
\put(147,125){.}
\put(148,128){.}
\put(149,131){.}
\put(150,133){.}
\put(152,136){.}
\put(153,139){.}
\put(154,142){.}
\put(155,145){.}
\put(156,148){.}
\put(158,151){.}
\put(159,155){.}
\put(160,158){.}
\put(161,161){.}
\put(162,164){.}
\put(164,167){.}
\put(165,171){.}
\put(166,174){.}
\put(167,177){.}
\put(168,181){.}
\put(170,184){.}
\put(171,187){.}
\put(172,191){.}
\put(173,194){.}
\put(174,197){.}
\put(176,200){.}
\put(177,203){.}
\put(178,207){.}
\put(179,210){.}
\put(180,212){.}
\put(181,215){.}
\put(183,218){.}
\put(184,221){.}
\put(185,223){.}
\put(186,226){.}
\put(187,228){.}
\put(189,230){.}
\put(190,232){.}
\put(191,234){.}
\put(192,235){.}
\put(193,236){.}
\put(195,238){.}
\put(196,239){.}
\put(197,239){.}
\put(198,240){.}
\put(199,240){.}
\put(201,240){.}
\put(202,240){.}
\put(203,239){.}
\put(204,238){.}
\put(205,237){.}
\put(207,236){.}
\put(208,234){.}
\put(209,233){.}
\put(210,230){.}
\put(211,228){.}
\put(213,225){.}
\put(214,222){.}
\put(215,219){.}
\put(216,216){.}
\put(217,212){.}
\put(219,208){.}
\put(220,204){.}
\put(221,200){.}
\put(222,196){.}
\put(223,191){.}
\put(224,186){.}
\put(226,181){.}
\put(227,176){.}
\put(228,171){.}
\put(229,166){.}
\put(230,161){.}
\put(232,155){.}
\put(233,150){.}
\put(234,145){.}
\put(235,139){.}
\put(236,134){.}
\put(238,129){.}
\put(239,124){.}
\put(240,119){.}
\put(241,114){.}
\put(242,109){.}
\put(244,105){.}
\put(245,100){.}
\put(246,96){.}
\put(247,92){.}
\put(248,88){.}
\put(250,84){.}
\put(251,80){.}
\put(252,77){.}
\put(253,74){.}
\put(254,71){.}
\put(256,68){.}
\put(257,65){.}
\put(258,63){.}
\put(259,60){.}
\put(260,58){.}
\put(261,56){.}
\put(263,54){.}
\put(264,53){.}
\put(265,51){.}
\put(266,50){.}
\put(267,49){.}
\put(269,48){.}
\put(270,47){.}
\put(271,46){.}
\put(272,45){.}
\put(273,44){.}
\put(275,44){.}
\put(276,43){.}
\put(277,43){.}
\put(278,42){.}
\put(279,42){.}
\put(281,42){.}
\put(282,41){.}
\put(283,41){.}
\put(284,41){.}
\put(285,41){.}
\put(287,41){.}
\put(288,41){.}
\put(289,40){.}
\put(290,40){.}
\put(291,40){.}
\put(293,40){.}
\put(294,40){.}
\put(295,40){.}
\put(296,40){.}
\put(297,40){.}
\put(299,40){.}
\put(300,40){.}
\put(301,40){.}
\put(302,40){.}
\put(303,40){.}
\put(304,40){.}
\put(306,40){.}
\put(307,40){.}
\put(308,40){.}
\put(309,40){.}
\put(310,40){.}
\put(312,40){.}
\put(313,40){.}
\put(314,40){.}
\put(315,40){.}
\put(316,40){.}
\put(318,40){.}
\put(319,40){.}
\put(320,40){.}
\put(321,40){.}
\put(322,40){.}
\put(324,40){.}
\put(325,40){.}
\put(326,40){.}
\put(327,40){.}
\put(328,40){.}
\put(330,40){.}
\put(331,40){.}
\put(332,40){.}
\put(333,40){.}
\put(334,40){.}
\put(336,40){.}
\put(337,40){.}
\put(338,40){.}
\put(339,40){.}
\put(340,40){.}
\put(341,40){.}
\put(343,40){.}
\put(344,40){.}
\put(345,40){.}
\put(346,40){.}
\put(347,40){.}
\put(349,40){.}
\put(350,40){.}
\put(50,41){.}
\put(51,41){.}
\put(52,41){.}
\put(53,41){.}
\put(53,41){.}
\put(54,41){.}
\put(55,41){.}
\put(55,41){.}
\put(56,41){.}
\put(57,41){.}
\put(58,41){.}
\put(58,41){.}
\put(59,41){.}
\put(60,41){.}
\put(61,41){.}
\put(61,42){.}
\put(62,42){.}
\put(63,42){.}
\put(64,42){.}
\put(64,42){.}
\put(65,42){.}
\put(66,42){.}
\put(66,42){.}
\put(67,42){.}
\put(68,42){.}
\put(69,42){.}
\put(69,42){.}
\put(70,42){.}
\put(71,43){.}
\put(72,43){.}
\put(72,43){.}
\put(73,43){.}
\put(74,43){.}
\put(75,43){.}
\put(75,43){.}
\put(76,43){.}
\put(77,43){.}
\put(77,44){.}
\put(78,44){.}
\put(79,44){.}
\put(80,44){.}
\put(80,44){.}
\put(81,44){.}
\put(82,44){.}
\put(83,45){.}
\put(83,45){.}
\put(84,45){.}
\put(85,45){.}
\put(86,45){.}
\put(86,46){.}
\put(87,46){.}
\put(88,46){.}
\put(88,46){.}
\put(89,46){.}
\put(90,47){.}
\put(91,47){.}
\put(91,47){.}
\put(92,47){.}
\put(93,48){.}
\put(94,48){.}
\put(94,48){.}
\put(95,49){.}
\put(96,49){.}
\put(97,49){.}
\put(97,49){.}
\put(98,50){.}
\put(99,50){.}
\put(99,51){.}
\put(100,51){.}
\put(101,51){.}
\put(102,52){.}
\put(102,52){.}
\put(103,53){.}
\put(104,53){.}
\put(105,53){.}
\put(105,54){.}
\put(106,54){.}
\put(107,55){.}
\put(108,55){.}
\put(108,56){.}
\put(109,56){.}
\put(110,57){.}
\put(110,58){.}
\put(111,58){.}
\put(112,59){.}
\put(113,59){.}
\put(113,60){.}
\put(114,61){.}
\put(115,61){.}
\put(116,62){.}
\put(116,63){.}
\put(117,64){.}
\put(118,64){.}
\put(118,65){.}
\put(119,66){.}
\put(120,67){.}
\put(121,68){.}
\put(121,69){.}
\put(122,69){.}
\put(123,70){.}
\put(124,71){.}
\put(124,72){.}
\put(125,73){.}
\put(126,74){.}
\put(127,75){.}
\put(127,76){.}
\put(128,78){.}
\put(129,79){.}
\put(129,80){.}
\put(130,81){.}
\put(131,82){.}
\put(132,84){.}
\put(132,85){.}
\put(133,86){.}
\put(134,87){.}
\put(135,89){.}
\put(135,90){.}
\put(136,92){.}
\put(137,93){.}
\put(138,94){.}
\put(138,96){.}
\put(139,97){.}
\put(140,99){.}
\put(140,101){.}
\put(141,102){.}
\put(142,104){.}
\put(143,106){.}
\put(143,107){.}
\put(144,109){.}
\put(145,111){.}
\put(146,112){.}
\put(146,114){.}
\put(147,116){.}
\put(148,118){.}
\put(149,120){.}
\put(149,122){.}
\put(150,124){.}
\put(151,126){.}
\put(151,128){.}
\put(152,130){.}
\put(153,132){.}
\put(154,134){.}
\put(154,136){.}
\put(155,138){.}
\put(156,140){.}
\put(157,142){.}
\put(157,144){.}
\put(158,146){.}
\put(159,149){.}
\put(160,151){.}
\put(160,153){.}
\put(161,155){.}
\put(162,158){.}
\put(162,160){.}
\put(163,162){.}
\put(164,164){.}
\put(165,167){.}
\put(165,169){.}
\put(166,171){.}
\put(167,173){.}
\put(168,176){.}
\put(168,178){.}
\put(169,180){.}
\put(170,182){.}
\put(171,184){.}
\put(171,187){.}
\put(172,189){.}
\put(173,191){.}
\put(173,193){.}
\put(174,195){.}
\put(175,197){.}
\put(176,200){.}
\put(176,202){.}
\put(177,204){.}
\put(178,206){.}
\put(179,208){.}
\put(179,210){.}
\put(180,211){.}
\put(181,213){.}
\put(182,215){.}
\put(182,217){.}
\put(183,219){.}
\put(184,220){.}
\put(184,222){.}
\put(185,223){.}
\put(186,225){.}
\put(187,226){.}
\put(187,228){.}
\put(188,229){.}
\put(189,230){.}
\put(190,231){.}
\put(190,233){.}
\put(191,234){.}
\put(192,235){.}
\put(192,236){.}
\put(193,236){.}
\put(194,237){.}
\put(195,238){.}
\put(195,238){.}
\put(196,239){.}
\put(197,239){.}
\put(198,240){.}
\put(198,240){.}
\put(199,240){.}
\put(200,240){.}
\put(201,240){.}
\put(201,240){.}
\put(202,240){.}
\put(203,239){.}
\put(203,239){.}
\put(204,239){.}
\put(205,238){.}
\put(206,237){.}
\put(206,237){.}
\put(207,236){.}
\put(208,235){.}
\put(209,234){.}
\put(209,233){.}
\put(210,232){.}
\put(211,230){.}
\put(212,229){.}
\put(212,228){.}
\put(213,226){.}
\put(214,224){.}
\put(214,223){.}
\put(215,221){.}
\put(216,219){.}
\put(217,217){.}
\put(217,215){.}
\put(218,213){.}
\put(219,211){.}
\put(220,209){.}
\put(220,207){.}
\put(221,205){.}
\put(222,202){.}
\put(223,200){.}
\put(223,197){.}
\put(224,195){.}
\put(225,192){.}
\put(225,190){.}
\put(226,187){.}
\put(227,185){.}
\put(228,182){.}
\put(228,179){.}
\put(229,177){.}
\put(230,174){.}
\put(231,171){.}
\put(231,168){.}
\put(232,165){.}
\put(233,163){.}
\put(234,160){.}
\put(234,157){.}
\put(235,154){.}
\put(236,151){.}
\put(236,149){.}
\put(237,146){.}
\put(238,143){.}
\put(239,140){.}
\put(239,137){.}
\put(240,135){.}
\put(241,132){.}
\put(242,129){.}
\put(242,127){.}
\put(243,124){.}
\put(244,121){.}
\put(245,119){.}
\put(245,116){.}
\put(246,114){.}
\put(247,111){.}
\put(247,109){.}
\put(248,106){.}
\put(249,104){.}
\put(250,102){.}
\put(250,99){.}
\put(251,97){.}
\put(252,95){.}
\put(253,93){.}
\put(253,91){.}
\put(254,89){.}
\put(255,87){.}
\put(255,85){.}
\put(256,83){.}
\put(257,81){.}
\put(258,79){.}
\put(258,78){.}
\put(259,76){.}
\put(260,74){.}
\put(261,73){.}
\put(261,71){.}
\put(262,70){.}
\put(263,68){.}
\put(264,67){.}
\put(264,66){.}
\put(265,64){.}
\put(266,63){.}
\put(266,62){.}
\put(267,61){.}
\put(268,60){.}
\put(269,59){.}
\put(269,58){.}
\put(270,57){.}
\put(271,56){.}
\put(272,55){.}
\put(272,54){.}
\put(273,53){.}
\put(274,52){.}
\put(275,52){.}
\put(275,51){.}
\put(276,50){.}
\put(277,50){.}
\put(277,49){.}
\put(278,48){.}
\put(279,48){.}
\put(280,47){.}
\put(280,47){.}
\put(281,46){.}
\put(282,46){.}
\put(283,46){.}
\put(283,45){.}
\put(284,45){.}
\put(285,45){.}
\put(286,44){.}
\put(286,44){.}
\put(287,44){.}
\put(288,43){.}
\put(288,43){.}
\put(289,43){.}
\put(290,43){.}
\put(291,42){.}
\put(291,42){.}
\put(292,42){.}
\put(293,42){.}
\put(50,41){.}
\put(51,41){.}
\put(51,41){.}
\put(52,41){.}
\put(53,41){.}
\put(53,41){.}
\put(54,41){.}
\put(55,41){.}
\put(55,41){.}
\put(56,41){.}
\put(57,41){.}
\put(57,41){.}
\put(58,41){.}
\put(59,41){.}
\put(59,41){.}
\put(60,41){.}
\put(61,41){.}
\put(61,41){.}
\put(62,41){.}
\put(63,41){.}
\put(63,41){.}
\put(64,41){.}
\put(65,41){.}
\put(65,41){.}
\put(66,42){.}
\put(67,42){.}
\put(67,42){.}
\put(68,42){.}
\put(69,42){.}
\put(69,42){.}
\put(70,42){.}
\put(71,42){.}
\put(71,42){.}
\put(72,42){.}
\put(73,42){.}
\put(73,42){.}
\put(74,42){.}
\put(75,43){.}
\put(75,43){.}
\put(76,43){.}
\put(77,43){.}
\put(77,43){.}
\put(78,43){.}
\put(79,43){.}
\put(79,43){.}
\put(80,43){.}
\put(81,44){.}
\put(81,44){.}
\put(82,44){.}
\put(83,44){.}
\put(83,44){.}
\put(84,44){.}
\put(84,44){.}
\put(85,45){.}
\put(86,45){.}
\put(86,45){.}
\put(87,45){.}
\put(88,45){.}
\put(88,45){.}
\put(89,46){.}
\put(90,46){.}
\put(90,46){.}
\put(91,46){.}
\put(92,46){.}
\put(92,47){.}
\put(93,47){.}
\put(94,47){.}
\put(94,47){.}
\put(95,48){.}
\put(96,48){.}
\put(96,48){.}
\put(97,48){.}
\put(98,49){.}
\put(98,49){.}
\put(99,49){.}
\put(100,50){.}
\put(100,50){.}
\put(101,50){.}
\put(102,51){.}
\put(102,51){.}
\put(103,51){.}
\put(104,52){.}
\put(104,52){.}
\put(105,53){.}
\put(106,53){.}
\put(106,53){.}
\put(107,54){.}
\put(108,54){.}
\put(108,55){.}
\put(109,55){.}
\put(110,56){.}
\put(110,56){.}
\put(111,57){.}
\put(112,57){.}
\put(112,58){.}
\put(113,58){.}
\put(114,59){.}
\put(114,60){.}
\put(115,60){.}
\put(116,61){.}
\put(116,62){.}
\put(117,62){.}
\put(117,63){.}
\put(118,64){.}
\put(119,64){.}
\put(119,65){.}
\put(120,66){.}
\put(121,67){.}
\put(121,67){.}
\put(122,68){.}
\put(123,69){.}
\put(123,70){.}
\put(124,71){.}
\put(125,72){.}
\put(125,73){.}
\put(126,73){.}
\put(127,74){.}
\put(127,75){.}
\put(128,76){.}
\put(129,77){.}
\put(129,78){.}
\put(130,80){.}
\put(131,81){.}
\put(131,82){.}
\put(132,83){.}
\put(133,84){.}
\put(133,85){.}
\put(134,86){.}
\put(135,88){.}
\put(135,89){.}
\put(136,90){.}
\put(137,92){.}
\put(137,93){.}
\put(138,94){.}
\put(139,96){.}
\put(139,97){.}
\put(140,98){.}
\put(141,100){.}
\put(141,101){.}
\put(142,103){.}
\put(143,104){.}
\put(143,106){.}
\put(144,107){.}
\put(145,109){.}
\put(145,111){.}
\put(146,112){.}
\put(147,114){.}
\put(147,116){.}
\put(148,117){.}
\put(149,119){.}
\put(149,121){.}
\put(150,123){.}
\put(150,124){.}
\put(151,126){.}
\put(152,128){.}
\put(152,130){.}
\put(153,132){.}
\put(154,134){.}
\put(154,135){.}
\put(155,137){.}
\put(156,139){.}
\put(156,141){.}
\put(157,143){.}
\put(158,145){.}
\put(158,147){.}
\put(159,149){.}
\put(160,151){.}
\put(160,153){.}
\put(161,155){.}
\put(162,157){.}
\put(162,159){.}
\put(163,161){.}
\put(164,163){.}
\put(164,165){.}
\put(165,168){.}
\put(166,170){.}
\put(166,172){.}
\put(167,174){.}
\put(168,176){.}
\put(168,178){.}
\put(169,180){.}
\put(170,182){.}
\put(170,184){.}
\put(171,186){.}
\put(172,188){.}
\put(172,190){.}
\put(173,192){.}
\put(174,194){.}
\put(174,196){.}
\put(175,198){.}
\put(176,200){.}
\put(176,201){.}
\put(177,203){.}
\put(178,205){.}
\put(178,207){.}
\put(179,209){.}
\put(180,210){.}
\put(180,212){.}
\put(181,214){.}
\put(182,215){.}
\put(182,217){.}
\put(183,219){.}
\put(183,220){.}
\put(184,221){.}
\put(185,223){.}
\put(185,224){.}
\put(186,226){.}
\put(187,227){.}
\put(187,228){.}
\put(188,229){.}
\put(189,230){.}
\put(189,231){.}
\put(190,232){.}
\put(191,233){.}
\put(191,234){.}
\put(192,235){.}
\put(193,236){.}
\put(193,237){.}
\put(194,237){.}
\put(195,238){.}
\put(195,238){.}
\put(196,239){.}
\put(197,239){.}
\put(197,239){.}
\put(198,240){.}
\put(199,240){.}
\put(199,240){.}
\put(200,240){.}
\put(201,240){.}
\put(201,240){.}
\put(202,240){.}
\put(203,239){.}
\put(203,239){.}
\put(204,239){.}
\put(205,238){.}
\put(205,238){.}
\put(206,237){.}
\put(207,236){.}
\put(207,236){.}
\put(208,235){.}
\put(209,234){.}
\put(209,233){.}
\put(210,232){.}
\put(211,231){.}
\put(211,230){.}
\put(212,228){.}
\put(213,227){.}
\put(213,226){.}
\put(214,224){.}
\put(215,223){.}
\put(215,221){.}
\put(216,220){.}
\put(217,218){.}
\put(217,216){.}
\put(218,214){.}
\put(218,213){.}
\put(219,211){.}
\put(220,209){.}
\put(220,207){.}
\put(221,205){.}
\put(222,203){.}
\put(222,201){.}
\put(223,198){.}
\put(224,196){.}
\put(224,194){.}
\put(225,192){.}
\put(226,189){.}
\put(226,187){.}
\put(227,185){.}
\put(228,182){.}
\put(228,180){.}
\put(229,178){.}
\put(230,175){.}
\put(230,173){.}
\put(231,170){.}
\put(232,168){.}
\put(232,165){.}
\put(233,163){.}
\put(234,160){.}
\put(234,158){.}
\put(235,155){.}
\put(236,153){.}
\put(236,150){.}
\put(237,148){.}
\put(238,145){.}
\put(238,143){.}
\put(239,140){.}
\put(240,138){.}
\put(240,135){.}
\put(241,133){.}
\put(242,131){.}
\put(242,128){.}
\put(243,126){.}
\put(244,124){.}
\put(244,121){.}
\put(245,119){.}
\put(246,117){.}
\put(246,114){.}
\put(247,112){.}
\put(248,110){.}
\put(248,108){.}
\put(249,106){.}
\put(250,104){.}
\put(250,102){.}
\put(251,100){.}
\put(251,98){.}
\put(252,96){.}
\put(253,94){.}
\put(253,92){.}
\put(254,90){.}
\put(255,88){.}
\put(255,87){.}
\put(256,85){.}
\put(257,83){.}
\put(257,82){.}
\put(258,80){.}
\put(259,78){.}
\put(259,77){.}
\put(260,75){.}
\put(261,74){.}
\put(261,73){.}
\put(262,71){.}
\put(263,70){.}
\put(263,69){.}
\put(264,68){.}
\put(265,66){.}
\put(265,65){.}
\put(266,64){.}
\put(267,63){.}
\put(267,62){.}
\put(268,61){.}
\put(269,60){.}
\put(269,59){.}
\put(270,58){.}
\put(271,57){.}
\put(271,56){.}
\put(272,56){.}
\put(273,55){.}
\put(273,54){.}
\put(274,53){.}
\put(275,53){.}
\put(275,52){.}
\put(276,51){.}
\put(277,51){.}
\put(51,40){.}
\put(52,40){.}
\put(53,40){.}
\put(54,40){.}
\put(55,40){.}
\put(56,40){.}
\put(57,40){.}
\put(58,40){.}
\put(59,40){.}
\put(60,40){.}
\put(61,40){.}
\put(62,40){.}
\put(63,40){.}
\put(64,40){.}
\put(65,40){.}
\put(66,40){.}
\put(67,40){.}
\put(68,40){.}
\put(69,40){.}
\put(70,40){.}
\put(71,40){.}
\put(72,40){.}
\put(73,40){.}
\put(74,40){.}
\put(75,40){.}
\put(76,40){.}
\put(77,40){.}
\put(78,40){.}
\put(79,41){.}
\put(80,41){.}
\put(81,41){.}
\put(82,41){.}
\put(83,41){.}
\put(84,41){.}
\put(85,41){.}
\put(86,41){.}
\put(87,41){.}
\put(88,41){.}
\put(89,41){.}
\put(90,41){.}
\put(91,42){.}
\put(92,42){.}
\put(93,42){.}
\put(94,42){.}
\put(95,42){.}
\put(96,42){.}
\put(97,43){.}
\put(98,43){.}
\put(99,43){.}
\put(100,43){.}
\put(101,44){.}
\put(102,44){.}
\put(103,44){.}
\put(104,45){.}
\put(105,45){.}
\put(106,45){.}
\put(107,46){.}
\put(108,46){.}
\put(109,47){.}
\put(110,47){.}
\put(111,48){.}
\put(112,48){.}
\put(113,49){.}
\put(114,50){.}
\put(115,50){.}
\put(116,51){.}
\put(117,52){.}
\put(118,53){.}
\put(119,54){.}
\put(120,54){.}
\put(121,55){.}
\put(122,56){.}
\put(123,58){.}
\put(124,59){.}
\put(125,60){.}
\put(126,61){.}
\put(127,62){.}
\put(128,64){.}
\put(129,65){.}
\put(130,67){.}
\put(131,68){.}
\put(132,70){.}
\put(133,71){.}
\put(134,73){.}
\put(135,75){.}
\put(136,77){.}
\put(137,79){.}
\put(138,81){.}
\put(139,83){.}
\put(140,85){.}
\put(141,87){.}
\put(142,90){.}
\put(143,92){.}
\put(144,95){.}
\put(145,97){.}
\put(146,100){.}
\put(147,102){.}
\put(148,105){.}
\put(149,108){.}
\put(150,111){.}
\put(151,114){.}
\put(152,117){.}
\put(153,120){.}
\put(154,123){.}
\put(155,126){.}
\put(156,129){.}
\put(157,132){.}
\put(158,135){.}
\put(159,139){.}
\put(160,142){.}
\put(161,145){.}
\put(162,149){.}
\put(163,152){.}
\put(164,156){.}
\put(165,159){.}
\put(166,163){.}
\put(167,166){.}
\put(168,169){.}
\put(169,173){.}
\put(170,176){.}
\put(171,180){.}
\put(172,183){.}
\put(173,186){.}
\put(174,189){.}
\put(175,193){.}
\put(176,196){.}
\put(177,199){.}
\put(178,202){.}
\put(179,205){.}
\put(180,208){.}
\put(181,210){.}
\put(182,213){.}
\put(183,216){.}
\put(184,218){.}
\put(185,221){.}
\put(186,223){.}
\put(187,225){.}
\put(188,227){.}
\put(189,229){.}
\put(190,231){.}
\put(191,232){.}
\put(192,234){.}
\put(193,235){.}
\put(194,236){.}
\put(195,237){.}
\put(196,238){.}
\put(197,239){.}
\put(198,239){.}
\put(199,240){.}
\put(200,240){.}
\put(201,240){.}
\put(202,240){.}
\put(203,240){.}
\put(204,239){.}
\put(205,239){.}
\put(206,238){.}
\put(207,237){.}
\put(208,236){.}
\put(209,235){.}
\put(210,234){.}
\put(211,232){.}
\put(212,231){.}
\put(213,229){.}
\put(214,227){.}
\put(215,225){.}
\put(216,223){.}
\put(217,221){.}
\put(218,218){.}
\put(219,216){.}
\put(220,213){.}
\put(221,210){.}
\put(222,208){.}
\put(223,205){.}
\put(224,202){.}
\put(225,199){.}
\put(226,196){.}
\put(227,193){.}
\put(228,189){.}
\put(229,186){.}
\put(230,183){.}
\put(231,180){.}
\put(232,176){.}
\put(233,173){.}
\put(234,169){.}
\put(235,166){.}
\put(236,163){.}
\put(237,159){.}
\put(238,156){.}
\put(239,152){.}
\put(240,149){.}
\put(241,145){.}
\put(242,142){.}
\put(243,139){.}
\put(244,135){.}
\put(245,132){.}
\put(246,129){.}
\put(247,126){.}
\put(248,123){.}
\put(249,120){.}
\put(250,117){.}
\put(251,114){.}
\put(252,111){.}
\put(253,108){.}
\put(254,105){.}
\put(255,102){.}
\put(256,100){.}
\put(257,97){.}
\put(258,95){.}
\put(259,92){.}
\put(260,90){.}
\put(261,87){.}
\put(262,85){.}
\put(263,83){.}
\put(264,81){.}
\put(265,79){.}
\put(266,77){.}
\put(267,75){.}
\put(268,73){.}
\put(269,71){.}
\put(270,70){.}
\put(271,68){.}
\put(272,67){.}
\put(273,65){.}
\put(274,64){.}
\put(275,62){.}
\put(276,61){.}
\put(277,60){.}
\put(278,59){.}
\put(279,58){.}
\put(280,56){.}
\put(281,55){.}
\put(282,54){.}
\put(283,54){.}
\put(284,53){.}
\put(285,52){.}
\put(286,51){.}
\put(287,50){.}
\put(288,50){.}
\put(289,49){.}
\put(290,48){.}
\put(291,48){.}
\put(292,47){.}
\put(293,47){.}
\put(294,46){.}
\put(295,46){.}
\put(296,45){.}
\put(297,45){.}
\put(298,45){.}
\put(299,44){.}
\put(300,44){.}
\put(301,44){.}
\put(302,43){.}
\put(303,43){.}
\put(304,43){.}
\put(305,43){.}
\put(306,42){.}
\put(307,42){.}
\put(308,42){.}
\put(309,42){.}
\put(310,42){.}
\put(311,42){.}
\put(312,41){.}
\put(313,41){.}
\put(314,41){.}
\put(315,41){.}
\put(316,41){.}
\put(317,41){.}
\put(318,41){.}
\put(319,41){.}
\put(320,41){.}
\put(321,41){.}
\put(322,41){.}
\put(323,41){.}
\put(324,40){.}
\put(325,40){.}
\put(326,40){.}
\put(327,40){.}
\put(328,40){.}
\put(329,40){.}
\put(330,40){.}
\put(331,40){.}
\put(332,40){.}
\put(333,40){.}
\put(334,40){.}
\put(335,40){.}
\put(336,40){.}
\put(337,40){.}
\put(338,40){.}
\put(339,40){.}
\put(340,40){.}
\put(341,40){.}
\put(342,40){.}
\put(343,40){.}
\put(344,40){.}
\put(345,40){.}
\put(346,40){.}
\put(347,40){.}
\put(348,40){.}
\put(349,40){.}
\put(350,40){.}
\put(51,45){.}
\put(52,46){.}
\put(53,46){.}
\put(54,46){.}
\put(55,46){.}
\put(56,46){.}
\put(57,46){.}
\put(58,47){.}
\put(59,47){.}
\put(60,47){.}
\put(61,47){.}
\put(62,48){.}
\put(63,48){.}
\put(64,48){.}
\put(65,48){.}
\put(66,49){.}
\put(67,49){.}
\put(68,49){.}
\put(69,49){.}
\put(70,50){.}
\put(71,50){.}
\put(72,50){.}
\put(73,51){.}
\put(74,51){.}
\put(75,51){.}
\put(76,52){.}
\put(77,52){.}
\put(78,52){.}
\put(79,53){.}
\put(80,53){.}
\put(81,53){.}
\put(82,54){.}
\put(83,54){.}
\put(84,55){.}
\put(85,55){.}
\put(86,56){.}
\put(87,56){.}
\put(88,56){.}
\put(89,57){.}
\put(90,57){.}
\put(91,58){.}
\put(92,59){.}
\put(93,59){.}
\put(94,60){.}
\put(95,60){.}
\put(96,61){.}
\put(97,61){.}
\put(98,62){.}
\put(99,63){.}
\put(100,63){.}
\put(101,64){.}
\put(102,65){.}
\put(103,66){.}
\put(104,66){.}
\put(105,67){.}
\put(106,68){.}
\put(107,69){.}
\put(108,70){.}
\put(109,70){.}
\put(110,71){.}
\put(111,72){.}
\put(112,73){.}
\put(113,74){.}
\put(114,75){.}
\put(115,76){.}
\put(116,77){.}
\put(117,78){.}
\put(118,79){.}
\put(119,81){.}
\put(120,82){.}
\put(121,83){.}
\put(122,84){.}
\put(123,85){.}
\put(124,87){.}
\put(125,88){.}
\put(126,89){.}
\put(127,91){.}
\put(128,92){.}
\put(129,94){.}
\put(130,95){.}
\put(131,97){.}
\put(132,98){.}
\put(133,100){.}
\put(134,101){.}
\put(135,103){.}
\put(136,105){.}
\put(137,106){.}
\put(138,108){.}
\put(139,110){.}
\put(140,112){.}
\put(141,114){.}
\put(142,116){.}
\put(143,118){.}
\put(144,120){.}
\put(145,122){.}
\put(146,124){.}
\put(147,126){.}
\put(148,128){.}
\put(149,130){.}
\put(150,132){.}
\put(151,135){.}
\put(152,137){.}
\put(153,139){.}
\put(154,142){.}
\put(155,144){.}
\put(156,146){.}
\put(157,149){.}
\put(158,151){.}
\put(159,154){.}
\put(160,156){.}
\put(161,159){.}
\put(162,161){.}
\put(163,164){.}
\put(164,167){.}
\put(165,169){.}
\put(166,172){.}
\put(167,175){.}
\put(168,177){.}
\put(169,180){.}
\put(170,183){.}
\put(171,185){.}
\put(172,188){.}
\put(173,191){.}
\put(174,193){.}
\put(175,196){.}
\put(176,199){.}
\put(177,201){.}
\put(178,204){.}
\put(179,206){.}
\put(180,209){.}
\put(181,211){.}
\put(182,214){.}
\put(183,216){.}
\put(184,218){.}
\put(185,220){.}
\put(186,223){.}
\put(187,225){.}
\put(188,226){.}
\put(189,228){.}
\put(190,230){.}
\put(191,232){.}
\put(192,233){.}
\put(193,235){.}
\put(194,236){.}
\put(195,237){.}
\put(196,238){.}
\put(197,239){.}
\put(198,239){.}
\put(199,240){.}
\put(200,240){.}
\put(201,240){.}
\put(202,240){.}
\put(203,240){.}
\put(204,239){.}
\put(205,238){.}
\put(206,238){.}
\put(207,236){.}
\put(208,235){.}
\put(209,234){.}
\put(210,232){.}
\put(211,230){.}
\put(212,228){.}
\put(213,225){.}
\put(214,223){.}
\put(215,220){.}
\put(216,217){.}
\put(217,213){.}
\put(218,210){.}
\put(219,206){.}
\put(220,202){.}
\put(221,198){.}
\put(222,194){.}
\put(223,190){.}
\put(224,185){.}
\put(225,181){.}
\put(226,176){.}
\put(227,171){.}
\put(228,166){.}
\put(229,161){.}
\put(230,156){.}
\put(231,151){.}
\put(232,146){.}
\put(233,141){.}
\put(234,135){.}
\put(235,130){.}
\put(236,125){.}
\put(237,120){.}
\put(238,115){.}
\put(239,111){.}
\put(240,106){.}
\put(241,101){.}
\put(242,97){.}
\put(243,92){.}
\put(244,88){.}
\put(245,84){.}
\put(246,81){.}
\put(247,77){.}
\put(248,74){.}
\put(249,70){.}
\put(250,67){.}
\put(251,64){.}
\put(252,62){.}
\put(253,59){.}
\put(254,57){.}
\put(255,55){.}
\put(256,53){.}
\put(257,52){.}
\put(258,50){.}
\put(259,49){.}
\put(260,47){.}
\put(261,46){.}
\put(262,45){.}
\put(263,44){.}
\put(264,44){.}
\put(265,43){.}
\put(266,43){.}
\put(267,42){.}
\put(268,42){.}
\put(269,41){.}
\put(270,41){.}
\put(271,41){.}
\put(272,41){.}
\put(273,41){.}
\put(274,40){.}
\put(275,40){.}
\put(276,40){.}
\put(277,40){.}
\put(278,40){.}
\put(279,40){.}
\put(280,40){.}
\put(281,40){.}
\put(282,40){.}
\put(283,40){.}
\put(284,40){.}
\put(285,40){.}
\put(286,40){.}
\put(287,40){.}
\put(288,40){.}
\put(289,40){.}
\put(290,40){.}
\put(291,40){.}
\put(292,40){.}
\put(293,40){.}
\put(294,40){.}
\put(295,40){.}
\put(296,40){.}
\put(297,40){.}
\put(298,40){.}
\put(25,40){\vector(1,0){400}}
\put(200,25){0}
\put(250,25){1}
\put(170,5){$Figure\ 1$}
\end{picture}

$$Form \ of \ spectrum \ under \ the \ different \ powers \ of \ deformation.$$

\pagebreak

\begin{picture}(350,470)
\put(65,320){.}
\put(66,321){.}
\put(67,322){.}
\put(68,322){.}
\put(69,323){.}
\put(70,324){.}
\put(71,324){.}
\put(72,325){.}
\put(73,326){.}
\put(74,326){.}
\put(75,327){.}
\put(76,328){.}
\put(77,328){.}
\put(78,329){.}
\put(79,330){.}
\put(80,330){.}
\put(81,331){.}
\put(82,332){.}
\put(83,332){.}
\put(84,333){.}
\put(85,333){.}
\put(86,334){.}
\put(87,335){.}
\put(88,335){.}
\put(89,336){.}
\put(90,336){.}
\put(91,337){.}
\put(92,337){.}
\put(93,338){.}
\put(94,338){.}
\put(95,339){.}
\put(96,339){.}
\put(97,340){.}
\put(98,340){.}
\put(99,341){.}
\put(100,341){.}
\put(101,342){.}
\put(102,342){.}
\put(103,342){.}
\put(104,343){.}
\put(105,343){.}
\put(106,344){.}
\put(107,344){.}
\put(108,344){.}
\put(109,345){.}
\put(110,345){.}
\put(111,345){.}
\put(112,346){.}
\put(113,346){.}
\put(114,346){.}
\put(115,347){.}
\put(116,347){.}
\put(117,347){.}
\put(118,347){.}
\put(119,348){.}
\put(120,348){.}
\put(121,348){.}
\put(122,348){.}
\put(123,348){.}
\put(124,349){.}
\put(125,349){.}
\put(126,349){.}
\put(127,349){.}
\put(128,349){.}
\put(129,349){.}
\put(130,349){.}
\put(131,349){.}
\put(132,349){.}
\put(133,349){.}
\put(134,349){.}
\put(135,349){.}
\put(136,349){.}
\put(137,349){.}
\put(138,349){.}
\put(139,349){.}
\put(140,349){.}
\put(141,349){.}
\put(142,349){.}
\put(143,348){.}
\put(144,348){.}
\put(145,348){.}
\put(146,348){.}
\put(147,347){.}
\put(148,347){.}
\put(149,347){.}
\put(150,347){.}
\put(151,346){.}
\put(152,346){.}
\put(153,345){.}
\put(154,345){.}
\put(155,345){.}
\put(156,344){.}
\put(157,344){.}
\put(158,343){.}
\put(159,343){.}
\put(160,342){.}
\put(161,341){.}
\put(162,341){.}
\put(163,340){.}
\put(164,339){.}
\put(165,339){.}
\put(166,338){.}
\put(167,337){.}
\put(168,337){.}
\put(169,336){.}
\put(170,335){.}
\put(171,334){.}
\put(172,333){.}
\put(173,332){.}
\put(174,331){.}
\put(175,330){.}
\put(176,329){.}
\put(177,328){.}
\put(178,327){.}
\put(179,326){.}
\put(180,325){.}
\put(181,324){.}
\put(182,323){.}
\put(183,321){.}
\put(184,320){.}
\put(50,44){.}
\put(51,44){.}
\put(53,44){.}
\put(54,44){.}
\put(55,44){.}
\put(56,44){.}
\put(58,45){.}
\put(59,45){.}
\put(60,45){.}
\put(61,45){.}
\put(63,45){.}
\put(64,46){.}
\put(65,46){.}
\put(66,46){.}
\put(68,46){.}
\put(69,47){.}
\put(70,47){.}
\put(71,47){.}
\put(73,47){.}
\put(74,48){.}
\put(75,48){.}
\put(76,48){.}
\put(78,49){.}
\put(79,49){.}
\put(80,49){.}
\put(82,50){.}
\put(83,50){.}
\put(84,51){.}
\put(85,51){.}
\put(87,51){.}
\put(88,52){.}
\put(89,52){.}
\put(90,53){.}
\put(92,53){.}
\put(93,54){.}
\put(94,54){.}
\put(95,55){.}
\put(97,56){.}
\put(98,56){.}
\put(99,57){.}
\put(100,57){.}
\put(102,58){.}
\put(103,59){.}
\put(104,60){.}
\put(105,60){.}
\put(107,61){.}
\put(108,62){.}
\put(109,63){.}
\put(110,64){.}
\put(112,64){.}
\put(113,65){.}
\put(114,66){.}
\put(115,67){.}
\put(117,68){.}
\put(118,69){.}
\put(119,71){.}
\put(120,72){.}
\put(122,73){.}
\put(123,74){.}
\put(124,75){.}
\put(125,77){.}
\put(127,78){.}
\put(128,79){.}
\put(129,81){.}
\put(130,82){.}
\put(132,84){.}
\put(133,85){.}
\put(134,87){.}
\put(135,88){.}
\put(137,90){.}
\put(138,92){.}
\put(139,94){.}
\put(140,95){.}
\put(142,97){.}
\put(143,99){.}
\put(144,101){.}
\put(145,103){.}
\put(147,105){.}
\put(148,107){.}
\put(149,109){.}
\put(150,112){.}
\put(152,114){.}
\put(153,116){.}
\put(154,118){.}
\put(155,121){.}
\put(157,123){.}
\put(158,126){.}
\put(159,128){.}
\put(161,130){.}
\put(162,133){.}
\put(163,135){.}
\put(164,138){.}
\put(166,141){.}
\put(167,143){.}
\put(168,146){.}
\put(169,148){.}
\put(171,151){.}
\put(172,153){.}
\put(173,156){.}
\put(174,158){.}
\put(176,161){.}
\put(177,163){.}
\put(178,166){.}
\put(179,168){.}
\put(181,170){.}
\put(182,172){.}
\put(183,174){.}
\put(184,176){.}
\put(186,178){.}
\put(187,180){.}
\put(188,182){.}
\put(189,183){.}
\put(191,185){.}
\put(192,186){.}
\put(193,187){.}
\put(194,188){.}
\put(196,189){.}
\put(197,189){.}
\put(198,190){.}
\put(199,190){.}
\put(201,190){.}
\put(202,190){.}
\put(203,189){.}
\put(204,189){.}
\put(206,188){.}
\put(207,187){.}
\put(208,185){.}
\put(209,184){.}
\put(211,182){.}
\put(212,180){.}
\put(213,178){.}
\put(214,175){.}
\put(216,172){.}
\put(217,169){.}
\put(218,166){.}
\put(219,163){.}
\put(221,160){.}
\put(222,156){.}
\put(223,152){.}
\put(224,148){.}
\put(226,144){.}
\put(227,140){.}
\put(228,136){.}
\put(229,132){.}
\put(231,127){.}
\put(232,123){.}
\put(233,119){.}
\put(234,114){.}
\put(236,110){.}
\put(237,106){.}
\put(238,102){.}
\put(239,98){.}
\put(241,94){.}
\put(242,90){.}
\put(243,86){.}
\put(245,83){.}
\put(246,79){.}
\put(247,76){.}
\put(248,73){.}
\put(250,70){.}
\put(251,67){.}
\put(252,64){.}
\put(253,62){.}
\put(255,60){.}
\put(256,58){.}
\put(257,56){.}
\put(258,54){.}
\put(260,52){.}
\put(261,51){.}
\put(262,49){.}
\put(263,48){.}
\put(265,47){.}
\put(266,46){.}
\put(267,45){.}
\put(268,44){.}
\put(270,44){.}
\put(271,43){.}
\put(272,43){.}
\put(273,42){.}
\put(275,42){.}
\put(276,42){.}
\put(277,41){.}
\put(278,41){.}
\put(280,41){.}
\put(281,41){.}
\put(282,41){.}
\put(283,40){.}
\put(285,40){.}
\put(286,40){.}
\put(287,40){.}
\put(288,40){.}
\put(290,40){.}
\put(291,40){.}
\put(292,40){.}
\put(293,40){.}
\put(295,40){.}
\put(296,40){.}
\put(297,40){.}
\put(298,40){.}
\put(300,40){.}
\put(301,40){.}
\put(302,40){.}
\put(303,40){.}
\put(305,40){.}
\put(306,40){.}
\put(307,40){.}
\put(308,40){.}
\put(310,40){.}
\put(311,40){.}
\put(312,40){.}
\put(313,40){.}
\put(315,40){.}
\put(316,40){.}
\put(317,40){.}
\put(318,40){.}
\put(320,40){.}
\put(321,40){.}
\put(322,40){.}
\put(324,40){.}
\put(325,40){.}
\put(326,40){.}
\put(327,40){.}
\put(329,40){.}
\put(330,40){.}
\put(331,40){.}
\put(332,40){.}
\put(334,40){.}
\put(335,40){.}
\put(336,40){.}
\put(337,40){.}
\put(339,40){.}
\put(340,40){.}
\put(341,40){.}
\put(342,40){.}
\put(344,40){.}
\put(345,40){.}
\put(346,40){.}
\put(347,40){.}
\put(349,40){.}
\put(350,40){.}
\put(57,321){.}
\put(58,321){.}
\put(59,322){.}
\put(60,323){.}
\put(61,324){.}
\put(62,325){.}
\put(63,326){.}
\put(64,327){.}
\put(65,327){.}
\put(66,328){.}
\put(67,329){.}
\put(68,330){.}
\put(69,331){.}
\put(70,332){.}
\put(71,333){.}
\put(72,333){.}
\put(73,334){.}
\put(74,335){.}
\put(75,336){.}
\put(76,337){.}
\put(77,338){.}
\put(78,338){.}
\put(79,339){.}
\put(80,340){.}
\put(81,341){.}
\put(82,342){.}
\put(83,342){.}
\put(84,343){.}
\put(85,344){.}
\put(86,345){.}
\put(87,345){.}
\put(88,346){.}
\put(89,347){.}
\put(90,348){.}
\put(91,349){.}
\put(92,349){.}
\put(93,350){.}
\put(94,351){.}
\put(95,352){.}
\put(96,352){.}
\put(97,353){.}
\put(98,354){.}
\put(99,354){.}
\put(100,355){.}
\put(101,356){.}
\put(102,357){.}
\put(103,357){.}
\put(104,358){.}
\put(105,359){.}
\put(106,359){.}
\put(107,360){.}
\put(108,361){.}
\put(109,361){.}
\put(110,362){.}
\put(111,363){.}
\put(112,363){.}
\put(113,364){.}
\put(114,365){.}
\put(115,365){.}
\put(116,366){.}
\put(117,366){.}
\put(118,367){.}
\put(119,368){.}
\put(120,368){.}
\put(121,369){.}
\put(122,369){.}
\put(123,370){.}
\put(124,371){.}
\put(125,371){.}
\put(126,372){.}
\put(127,372){.}
\put(128,373){.}
\put(129,373){.}
\put(130,374){.}
\put(131,374){.}
\put(132,375){.}
\put(133,375){.}
\put(134,376){.}
\put(135,376){.}
\put(136,377){.}
\put(137,377){.}
\put(138,377){.}
\put(139,378){.}
\put(140,378){.}
\put(141,379){.}
\put(142,379){.}
\put(143,379){.}
\put(144,380){.}
\put(145,380){.}
\put(146,380){.}
\put(147,381){.}
\put(148,381){.}
\put(149,381){.}
\put(150,382){.}
\put(151,382){.}
\put(152,382){.}
\put(153,382){.}
\put(154,383){.}
\put(155,383){.}
\put(156,383){.}
\put(157,383){.}
\put(158,383){.}
\put(159,384){.}
\put(160,384){.}
\put(161,384){.}
\put(162,384){.}
\put(163,384){.}
\put(164,384){.}
\put(165,384){.}
\put(166,384){.}
\put(167,384){.}
\put(168,384){.}
\put(169,384){.}
\put(170,384){.}
\put(171,384){.}
\put(172,384){.}
\put(173,384){.}
\put(174,384){.}
\put(175,384){.}
\put(176,384){.}
\put(177,384){.}
\put(178,384){.}
\put(179,383){.}
\put(180,383){.}
\put(181,383){.}
\put(182,383){.}
\put(183,382){.}
\put(184,382){.}
\put(185,382){.}
\put(186,382){.}
\put(187,381){.}
\put(188,381){.}
\put(189,380){.}
\put(190,380){.}
\put(191,380){.}
\put(192,379){.}
\put(193,379){.}
\put(194,378){.}
\put(195,378){.}
\put(196,377){.}
\put(197,376){.}
\put(198,376){.}
\put(199,375){.}
\put(200,375){.}
\put(201,374){.}
\put(202,373){.}
\put(203,372){.}
\put(204,372){.}
\put(205,371){.}
\put(206,370){.}
\put(207,369){.}
\put(208,368){.}
\put(209,367){.}
\put(210,366){.}
\put(211,365){.}
\put(212,364){.}
\put(213,363){.}
\put(214,362){.}
\put(215,361){.}
\put(216,360){.}
\put(217,359){.}
\put(218,358){.}
\put(219,357){.}
\put(220,355){.}
\put(221,354){.}
\put(222,353){.}
\put(223,351){.}
\put(224,350){.}
\put(225,348){.}
\put(226,347){.}
\put(227,346){.}
\put(228,344){.}
\put(229,342){.}
\put(230,341){.}
\put(231,339){.}
\put(232,337){.}
\put(233,336){.}
\put(234,334){.}
\put(235,332){.}
\put(236,330){.}
\put(237,328){.}
\put(238,327){.}
\put(239,325){.}
\put(240,323){.}
\put(241,321){.}
\put(51,44){.}
\put(52,44){.}
\put(53,44){.}
\put(54,44){.}
\put(56,44){.}
\put(57,44){.}
\put(58,45){.}
\put(59,45){.}
\put(61,45){.}
\put(62,45){.}
\put(63,45){.}
\put(64,46){.}
\put(66,46){.}
\put(67,46){.}
\put(68,46){.}
\put(69,46){.}
\put(70,47){.}
\put(72,47){.}
\put(73,47){.}
\put(74,48){.}
\put(75,48){.}
\put(77,48){.}
\put(78,49){.}
\put(79,49){.}
\put(80,49){.}
\put(82,50){.}
\put(83,50){.}
\put(84,50){.}
\put(85,51){.}
\put(87,51){.}
\put(88,52){.}
\put(89,52){.}
\put(90,53){.}
\put(92,53){.}
\put(93,54){.}
\put(94,54){.}
\put(95,55){.}
\put(97,55){.}
\put(98,56){.}
\put(99,56){.}
\put(100,57){.}
\put(102,58){.}
\put(103,58){.}
\put(104,59){.}
\put(105,60){.}
\put(107,61){.}
\put(108,61){.}
\put(109,62){.}
\put(110,63){.}
\put(111,64){.}
\put(113,65){.}
\put(114,66){.}
\put(115,67){.}
\put(116,68){.}
\put(118,69){.}
\put(119,70){.}
\put(120,71){.}
\put(121,72){.}
\put(123,73){.}
\put(124,75){.}
\put(125,76){.}
\put(126,77){.}
\put(128,78){.}
\put(129,80){.}
\put(130,81){.}
\put(131,83){.}
\put(133,84){.}
\put(134,86){.}
\put(135,87){.}
\put(136,89){.}
\put(138,91){.}
\put(139,93){.}
\put(140,94){.}
\put(141,96){.}
\put(143,98){.}
\put(144,100){.}
\put(145,102){.}
\put(146,104){.}
\put(148,106){.}
\put(149,108){.}
\put(150,110){.}
\put(151,112){.}
\put(152,115){.}
\put(154,117){.}
\put(155,119){.}
\put(156,122){.}
\put(157,124){.}
\put(159,126){.}
\put(160,129){.}
\put(161,131){.}
\put(162,134){.}
\put(164,136){.}
\put(165,139){.}
\put(166,141){.}
\put(167,144){.}
\put(169,147){.}
\put(170,149){.}
\put(171,152){.}
\put(172,154){.}
\put(174,157){.}
\put(175,159){.}
\put(176,162){.}
\put(177,164){.}
\put(179,166){.}
\put(180,169){.}
\put(181,171){.}
\put(182,173){.}
\put(184,175){.}
\put(185,177){.}
\put(186,179){.}
\put(187,181){.}
\put(189,182){.}
\put(190,184){.}
\put(191,185){.}
\put(192,186){.}
\put(193,187){.}
\put(195,188){.}
\put(196,189){.}
\put(197,190){.}
\put(198,190){.}
\put(200,190){.}
\put(201,190){.}
\put(202,190){.}
\put(203,189){.}
\put(205,188){.}
\put(206,188){.}
\put(207,186){.}
\put(208,185){.}
\put(210,183){.}
\put(211,182){.}
\put(212,179){.}
\put(213,177){.}
\put(215,175){.}
\put(216,172){.}
\put(217,169){.}
\put(218,166){.}
\put(220,163){.}
\put(221,159){.}
\put(222,156){.}
\put(223,152){.}
\put(225,148){.}
\put(226,144){.}
\put(227,140){.}
\put(228,136){.}
\put(230,132){.}
\put(231,128){.}
\put(232,123){.}
\put(233,119){.}
\put(234,115){.}
\put(236,111){.}
\put(237,106){.}
\put(238,102){.}
\put(239,98){.}
\put(241,94){.}
\put(242,90){.}
\put(243,87){.}
\put(244,83){.}
\put(246,80){.}
\put(247,77){.}
\put(248,73){.}
\put(249,70){.}
\put(251,68){.}
\put(252,65){.}
\put(253,63){.}
\put(254,60){.}
\put(256,58){.}
\put(257,56){.}
\put(258,54){.}
\put(259,53){.}
\put(261,51){.}
\put(262,50){.}
\put(263,49){.}
\put(264,47){.}
\put(266,46){.}
\put(267,46){.}
\put(268,45){.}
\put(269,44){.}
\put(270,43){.}
\put(272,43){.}
\put(273,42){.}
\put(274,42){.}
\put(275,42){.}
\put(277,41){.}
\put(278,41){.}
\put(279,41){.}
\put(280,41){.}
\put(282,41){.}
\put(283,41){.}
\put(284,40){.}
\put(285,40){.}
\put(287,40){.}
\put(288,40){.}
\put(289,40){.}
\put(290,40){.}
\put(292,40){.}
\put(293,40){.}
\put(294,40){.}
\put(295,40){.}
\put(297,40){.}
\put(298,40){.}
\put(299,40){.}
\put(300,40){.}
\put(302,40){.}
\put(303,40){.}
\put(304,40){.}
\put(305,40){.}
\put(307,40){.}
\put(308,40){.}
\put(309,40){.}
\put(310,40){.}
\put(311,40){.}
\put(313,40){.}
\put(314,40){.}
\put(315,40){.}
\put(316,40){.}
\put(318,40){.}
\put(319,40){.}
\put(320,40){.}
\put(321,40){.}
\put(323,40){.}
\put(324,40){.}
\put(325,40){.}
\put(326,40){.}
\put(328,40){.}
\put(329,40){.}
\put(330,40){.}
\put(331,40){.}
\put(333,40){.}
\put(334,40){.}
\put(335,40){.}
\put(336,40){.}
\put(338,40){.}
\put(339,40){.}
\put(340,40){.}
\put(341,40){.}
\put(343,40){.}
\put(344,40){.}
\put(345,40){.}
\put(346,40){.}
\put(348,40){.}
\put(349,40){.}
\put(25,370){\vector(1,0){300}}
\put(100,320){\vector(0,1){120}}
\put(105,380){0}
\put(150,360){1}
\put(320,360){$z$}
\put(105,420){1}
\put(105,450){$\delta$}
\put(150,300){$Figure \  2, \ part\ a$ }
\put(25,40){\vector(1,0){400}}
\put(250,25){1}
\put(200,25){0}
\put(350,25){$z$}
\put(150,10){$Figure \  2, \ part\ b$ }
\end{picture}

$$ Weak \ dependence \ of \ the \ spectrum \ on \ the \ base \ of \ approximations. $$


\begin{thebibliography}{99}
\bibitem{3}Zeldovitch, J.B., Journ. Exper. and Theor.Phys. (USSR) vol.24, p.749 (1942)
\bibitem{13}
Kurasov, V.B.
Description of homogeneous and heterogeneous
nucleation in dynamic conditions, Deponed in VINITI number
5147-B,
1.06.89, 50 p.
\bibitem{15}Lifshitz, I.M. and  Slyozov, V.V.,
Journal of Exper.Phys. (USSR) vol.35, p.479 (1958)
\bibitem{16}Lifshitz, I.M. and  Slyozov, V.V. J.Phys.Chem.Solids vol.19,  p.35
(1961)
\bibitem{17}Kuni, F.M.and  Grinin, A.P.,
Colloid.J. (USSR) vol.46, p.23 (1984)
\bibitem{18}Kuni, F.M.,
Colloid.J. (USSR) vol.46, p.674 (1984)
\bibitem{PhysRev}
V.Kurasov, Phys.Rev. E, vol.49, p.3948 (1994).
\bibitem{Novosib}
Grinin, A.P., Kuni, F.M., Kurasov, V.B.,
Heterogeneous nucleation in vapor flow, In Gadiak (ed.):
 Mechanics of unhomogeneous
systems, Novosibirsk, (1985), p.86
\bibitem{Sevdec}
Kurasov, V.,  VINITI (Russia) 2594-B95, 28 p.
\bibitem{Specdec}
Kurasov, V.,
 VINITI (Russia) 2589-B95,
23 p.
\bibitem{Sevdin}
Kurasov, V.,
 VINITI (Russia) 2591-B95, 21p.
\bibitem{specdin}
Kurasov V.,  VINITI (Russia) 2593-B95, 25p.
\bibitem{Twomey}
Twomey, S., Geofis. pura e appl., vol. 43, number 2, p.243 (1959)
\bibitem{aero1}
Kurasov V, Kinetics of aerosol formation. 1. Decay of metastable phase
on several types of heterogeneous centers (to be published)
\bibitem{aero2}
Kurasov V, Kinetics of aerosol formation. 2. Decay of metastable phase
on heterogeneous centers with continious activity (to be published)
\bibitem{aero3}
Kurasov V, Kinetics of aerosol formation.3. Heterogeneous condensation
on several types of centers in dynamic conditions (to be published)
\end{thebibliography}
\end{document}